\documentclass[10pt,journal,compsoc]{IEEEtran}
\usepackage{amsmath,amsfonts}
\usepackage{algorithmic}
\usepackage{algorithm}
\usepackage{array}
\ifCLASSOPTIONcompsoc
 \usepackage[caption=false,font=footnotesize,labelfont=sf,textfont=sf]{subfig}
\else
 \usepackage[caption=false,font=footnotesize]{subfig}
\fi
\usepackage{textcomp}
\usepackage{stfloats}
\usepackage{url}
\usepackage{verbatim}
\usepackage[pdftex]{graphicx}
\ifCLASSOPTIONcompsoc
  \usepackage[nocompress]{cite}
\else
  \usepackage{cite}
\fi

\usepackage{booktabs}
\usepackage{colortbl}
\usepackage[dvipsnames]{xcolor}
\usepackage{adjustbox}
\graphicspath{ {./images/} }

\definecolor{NavyBlue}{RGB}{19, 40, 190}
\definecolor{SecBg}{RGB}{255, 220, 150}
\definecolor{BestBg}{RGB}{255, 160, 160}

\usepackage{amssymb}    % for real number sign
\usepackage{xcolor}
\usepackage[colorlinks = true,
            linkcolor = purple,
            urlcolor  = blue,
            citecolor = blue,
            anchorcolor = black]{hyperref}	% Color links to references, figures, etc.
\usepackage{xpatch}
\usepackage{orcidlink} % If needed for ORCID icons

\newcommand\Alpha{\mathrel{A}}
\newcommand\Beta{\mathrel{B}}
\newcommand{\defeq}{\mathrel{\mathop:}=}

\begin{document}
\bstctlcite{IEEEexample:BSTcontrol}     % https://tex.stackexchange.com/a/164513

\title{
Intensity Field Decomposition for \\Tissue-Guided Neural Tomography
}
\author{%
Meng-Xun~Li, %
Jin-Gang~Yu, %
Yuan~Gao, %
Cui~Huang, %
and Gui-Song~Xia% <-this % stops a space
\IEEEcompsocitemizethanks{
\IEEEcompsocthanksitem{M.-X. Li is with the School of Stomatology and School of Computer Science, Wuhan University, Wuhan 430072, China}% <-this % stops a space
\IEEEcompsocthanksitem{Y. Gao is with Electronic Information School and School of Computer Science, Wuhan University, Wuhan 430072, China}%
\IEEEcompsocthanksitem{G.-S. Xia is with the School of Computer Science, Wuhan University, Wuhan 430072, China}%
\IEEEcompsocthanksitem{J.-G. Yu is with the School of Automation Science and Engineering, South China University of Technology, Guangzhou 510641, China, and also with Pazhou Laboratory, Guangzhou 510335, China.}%
\IEEEcompsocthanksitem{C. Huang is with the School of Stomatology, Wuhan University, Wuhan 430072, China}% <-this % stops a space
\IEEEcompsocthanksitem{Corresponding authors: C. Huang and G.-S. Xia}
\IEEEcompsocthanksitem{
The source code will be released at 
\href{https://github.com/menxli/cbct_tnt}{https://github.com/menxli/cbct\_tnt}
}
}
}

\markboth{Preprint}{}

\IEEEtitleabstractindextext{%
\begin{abstract}
Cone-beam computed tomography (CBCT) typically requires hundreds of X-ray projections, which raises concerns about radiation exposure. 
While sparse-view reconstruction reduces the exposure by using fewer projections, it struggles to achieve satisfactory image quality. 
To address this challenge, this article introduces a novel sparse-view CBCT reconstruction method, 
which empowers the neural field with human tissue regularization. 
Our approach, termed {\em tissue-guided neural tomography} (TNT), is motivated by the distinct intensity differences between bone and soft tissue in CBCT. 
Intuitively, separating these components may aid the learning process of the neural field. 
More precisely, TNT comprises a heterogeneous quadruple network and the corresponding training strategy.
The network represents the intensity field as a combination of soft and hard tissue components, along with their respective textures. 
We train the network with guidance from estimated tissue projections, enabling efficient learning of the desired patterns for the network heads. 
Extensive experiments demonstrate the proposed method significantly improves the sparse-view CBCT reconstruction with a limited number of projections ranging from 10 to 60. 
Our method achieves comparable reconstruction quality with 
fewer projections and faster convergence compared to state-of-the-art neural rendering based methods. 
\end{abstract}

}

\maketitle
\IEEEdisplaynontitleabstractindextext
\IEEEpeerreviewmaketitle

\IEEEraisesectionheading{\section{Introduction}}
\IEEEPARstart{C}{one}-beam computed-tomography (CBCT) can provide high-resolution volumetric images of hard tissue and becomes one of the most widely used imaging modalities in dentistry and oral maxillofacial surgery~\cite{kaasalainen2021cbctreview}. 
Despite numerous benefits, the techniques used by commercial CBCT machines often require hundreds of X-ray projections, 
which thus raises concerns about the associated radiation dose and its potential adverse effects~\cite{stratis2019cbctconcern}. 
To mitigate this issue, reducing the number of projections for CBCT reconstruction becomes a longstanding topic.

Classical analytical or iterative computed-tomography (CT) reconstruction approaches, such as Feldkamp-Davis-Kress (FDK) algorithm~\cite{feldkamp1984fdk} and {\em Simultaneous Algebraic Reconstruction Technique} (SART)~\cite{andersen1984sart} have been widely used in medical imaging due to their effectiveness in reconstruction with sufficient projections. 
However, these methods often struggle from sparse-view conditions. 
This inadequacy arises due to the ill-posed nature of sparse-view reconstruction, wherein there is insufficient measurement data to provide comprehensive information about the imaged objects.
Existing works incorporate prior knowledge to deal with this issue, 
\textit{e.g.} adding regularization terms into the optimization process or applying denoising techniques to the reconstructed image~\cite{arridge2019solving, zhang2018ctregreview}. 
With the rapid development of deep learning in computer vision, deep models have garnered considerable attention and extensive investigation into the medical image reconstruction problem~\cite{zhu2018imrec_nature, genzel2023dlmedrecrobust, wang2020dl, liu2023deepeit}.
Nevertheless, achieving end-to-end CBCT reconstruction devoid of physical constraints remains a challenging task for deep neural networks, which requires substantial volumes of training data and computational resources~\cite{wang2020dl}. Consequently, a majority of these endeavors continue to adhere to the traditional analytical or iterative reconstruction paradigm.

A noteworthy advancement in 3D reconstruction involves neural implicit scene representation and neural rendering~\cite{tewari2022neuralrenderreview}. 
Applying neural rendering based reconstruction to CT scans is undoubtedly a highly promising avenue since the neural field can naturally incorporate physical constraints into the deep learning reconstruction framework. 
We have witnessed some pioneer works on CBCT reconstruction with neural rendering~\cite{ruckert2022neat, zha2022naf}
and some attempts on sparse view conditions~\cite{fang2022snaf, abril2022mednerf, lin2023difnet, sun2023xrayct, liu2024svcbct}. 
However, many assumptions on improving sparse-view rendering quality for natural scenes do not hold for CT images, particularly in the case of geometric prior. %~\cite{}. 
For instance, the depth continuity~\cite{yu2022monosdf, wei2023nerfingmvs, wang2023sparsenerf} and visibility constraint~\cite{somraj2023vipnerf} which are widely used as regularization that can introduce substantial improvement, do not exist in CT images due to the penetrative nature of X-ray.
Similarly, the patch similarity~\cite{Michael2022RegNeRF} does not necessarily apply because of parallax, especially in maxillofacial regions where complex anatomical structures present.

Moreover, despite the advancements, existing works on neural rendering for sparse-view CBCT mainly build upon techniques for natural scenes or are designed for generic volumetric images where the intensity field is typically represented by a single neural network output.
However, we contend that capturing the diverse features of CBCT in an end-to-end manner poses a formidable challenge for a single neural network. 
When the number of projections is sufficient, this limitation can be mitigated by leveraging the information with complex neural network architectures. 
However, under sparse-view scenarios, relying on a single network with large capacity becomes a drawback by overfitting to the limited observations.

In this article, we aim to improve the sparse-view CBCT reconstruction with neural fields by addressing the aforementioned challenges. Our method is based on the following observations:
1. The soft and hard tissue differ both physiologically and radiologically, this dissimilarity is evident in their nature; 
Moreover, CBCT has a low radiometric resolution~\cite{pauwels2015technicalofcbct}, leading to less variation within the two tissue components. 
Therefore, we assume that the soft and hard tissue in CBCT exhibit distinct distributions and should not be treated as a homogeneous entity. 
2. We observed that a single X-ray image is a routinely used diagnostic tool in orthodontics, where a cephalometry is used to evaluate the shape of the skull~\cite{devereux2011cepImportant}, which indicates that the bony structures can be inferred from a single X-ray projection. 
We hypothesize that tissue shape can be predicted from projections and used as a geometric prior for the sparse-view reconstruction.
This is an analogy to the depth regularization in the 3D reconstruction of natural scenes, 
the accurate estimation of which greatly assists sparse-view rendering~\cite{deng2022depthnerf, yuan2023sparsenerfrgbd}.

Among the related researches for natural scenes, it is widely applied to decouple the shape and appearance, so as to offer better control of an object \cite{schwarz2020graf, jang2021codenerf, peng2023codebody}, or disassemble to enable efficient inference \cite{garbin2021fastnerf, wadhwani2022squeezenerf}. In addition, the network capacity may be limited in certain branches to regularize the selected feature \cite{zhang2020nerfpp}.  
Likewise, to leverage the aforementioned insights, we propose using a combination of neural fields, each specialized in modeling decoupled features of the soft and hard tissue, while regularizing the tissue shape and texture to impose prior to the reconstruction process. 

\begin{figure}[tb]
    \centering
    \includegraphics[width=.99\linewidth]{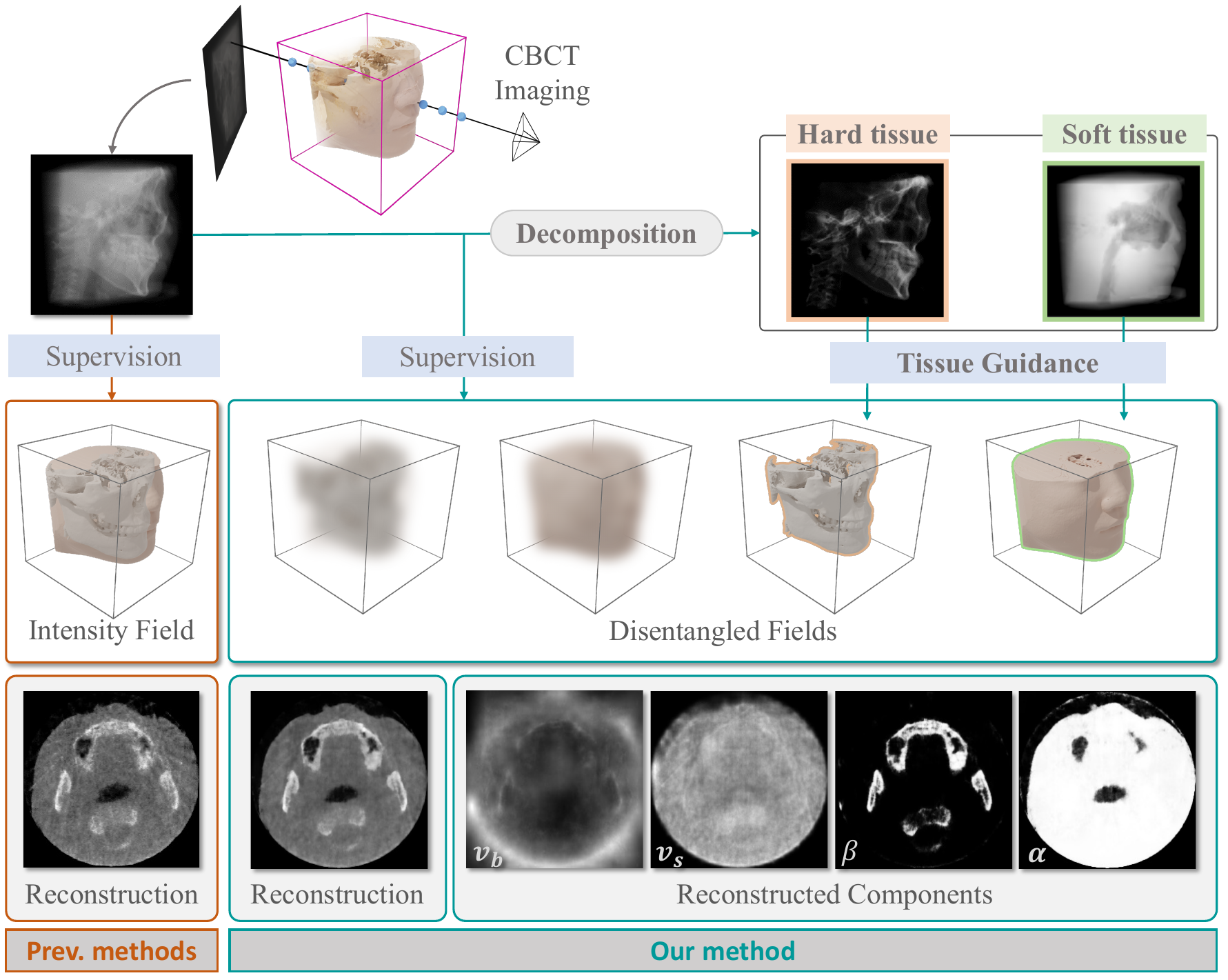}
    \caption{\small
    Overview of the proposed method. 
    Instead of directly regress a neural field to the projections, we introduce a novel approach by disentangling the field into 4 semantically meaningful components, which are then supervised with tissue projections that have clear clinical significance.
    }
    \label{fig:teaser}
\end{figure}

Specifically, we present a straightforward and intuitive strategy to segregate the volume density into soft and hard tissue shape and texture components. 
The proposed shapes are essentially the segmentation masks of the tissue, which are learned with the help of the predicted tissue projections (Fig.~\ref{fig:teaser}).
The textures are automatically adjusted to be smooth by the limited capacity of the neural network to provide limited variations among each tissue component.

Our approach achieves satisfying reconstruction quality of oral-maxillofacial region in varying degrees of sparsity from 10 to 60 projections, within a relatively low computational complexity. 
The main contributions of this work are two-fold: 
First, we propose a novel and meaningful intensity field decomposition that has direct clinical implications. The decomposition simplifies the learning process by modeling the diverse set of features present in CBCT with their specialized network while increasing the explainability and controllability of the neural field. 
Second, we leverage the tissue information from X-ray projection and incorporate tissue-specific guidance into the training process, which significantly accelerates convergence speed and enhances reconstruction quality.
% The source code will be released at 
% \href{https://github.com/menxli/cbct_tnt}{https://github.com/menxli/cbct\_tnt}.

\section{Related Work}

\subsection{Regularizing Neural Fields under Sparse Inputs}
Implicit 3D modeling has been utilized in the field of computer graphics for decades~\cite{Marschner2015fundamentalOfCG}. In recent years, neural networks have also been incorporated such as coordinate-based occupancy~\cite{mescheder2019occupancynetworks}, signed distance function~\cite{yu2022monosdf}, or feature representations~\cite{sitzmann2019srn}. 
One of the most well-studied methods among these works is Neural Radiance Fields (NeRF) and its extensions~\cite{mildenhall2021nerf}. These methods have achieved extraordinary success in view synthesis and a myriad of other 3D tasks. 

Despite the advances made in neural fields, 
their optimal performance heavily relies on substantial input images. 
The network tends to overfit to known views when the number of observations is limited, 
leading to artifacts. 
This issue can be further exacerbated with sparsely-coded inputs~\cite{muller2022instantngp, liu2020nsvf, chen2022tensorf}, 
which were designed to speed up learning 
and enhance the network's ability to capture fine-grained scene details. 

To address this issue, many studies incorporate regularization into the learning process. These include: 
\textit{Geometric constraints}: imposing surface depth~\cite{wei2023nerfingmvs, deng2022depthnerf}, normal~\cite{yu2022monosdf}, and visibility priors~\cite{somraj2023vipnerf}; 
\textit{Model capacity limitations}: applying frequency regularization~\cite{yang2023freenerf} and managing model size~\cite{zhang2020nerfpp, zhu2024mimlp}; 
\textit{Smoothness restrictions}: such as ray entropy~\cite{kim2022infonerf} and depth continuity~\cite{Michael2022RegNeRF, wang2023sparsenerf}; 
\textit{Data distribution priors}: incorporating patch distribution~\cite{Michael2022RegNeRF}, semantic similarity~\cite{jain2021dietnerf}, or diffusion-based gradient~\cite{wynn2023diffusionerf, zhou2023sparsefusion, deng2023nerdi, xu2023neurallift}. 
Notably, many of the above-mentioned restrictions are not directly applicable in the context of CT imaging because of the penetrative nature of X-ray. 
For example, depth constraints, which are commonly used as an auxiliary regularization~\cite{wei2023nerfingmvs, deng2022depthnerf, yu2022monosdf, Michael2022RegNeRF, wang2023sparsenerf, wynn2023diffusionerf, deng2023nerdi, xu2023neurallift}, 
do not apply to CT imaging due to the elimination of depth in projections. 
Similarly, visibility priors and patch similarity do not apply because of parallax. 
In addition, multi-modal semantic regularization  
used in~\cite{jain2021dietnerf, deng2023nerdi, xu2023neurallift}
may not transfer effectively, 
since the fine-grained internal structures of the human body are difficult to capture with a universal encoder under overlapped projections. 
To overcome these challenges, it is beneficial to explore domain-specific regularization strategies tailored for CT imaging.

% \vspace{-9pt}
\subsection{Sparse-view CT Reconstruction}
The presence of artifacts and noise are long-lasting challenges in CT reconstruction, particularly under sparse-view projections.
Classical reconstruction methods~\cite{feldkamp1984fdk, GORDON1970art} are hard to deal with these challenges since the sparse-view reconstruction itself is an ill-posed problem. 
Several approaches have been proposed to incorporate denoising regularization techniques into the process of classical reconstruction~\cite{Mahmoud2019improvedTV, Xu2020Schatten, zhang2018ctregreview}. 
However, these manually crafted methods are often suboptimal and may not perform well on sparse projections. 
Over the recent years, deep learning methods~\cite{Jiang2022dl,wang2020dl} have been extensively studied and have shown superior performance than manually designed regularizations. Examples include incorporating CNN into back-projection process~\cite{ye2018deep}, CNN denoising in different domains~\cite{CHAO2022536, zhang2018densenet, han2018framingunet, jin2017fbpconvnet}, using GANs for artifact reduction~\cite{liao2018gans}, or sinogram synthesis with neural networks~\cite{lee2019dnn}. 
The majority of these methods still heavily rely on the classical reconstruction theory and are constrained with inherent limitations. 
Notably, recent studies have explored novel avenues with the diffusion-based inverse problem solver~\cite{chung2023diffmbir, lee2023perpdiff}, leading to promising results. However, performing diffusion sampling on volume images is time-consuming. 

Recently, with the advancement in neural rendering, reconstruction using neural implicit scene representation has become another promising direction. 
These approaches provide a holistic representation and convert the discrete optimization over voxels into neural field learning with geometric constraints: NeAT\cite{ruckert2022neat} and NAF~\cite{zha2022naf} employed the octree-based optimization and hash encoding representation, respectively, to improve the training speed of CBCT reconstruction. 
MedNeRF\cite{abril2022mednerf} built a conditional generation network to synthesize new view projections under extremely sparse views. \cite{fang2022snaf} extended NAF~\cite{zha2022naf} with a deblurring network to optimize the projection in unknown views, improving volume reconstruction quality under sparse views.
Additionally, \cite{lin2023difnet, sun2023xrayct, liu2024svcbct} achieved CBCT reconstruction from extremely sparse view using cross-view feature fusion strategies.

Despite achieving impressive results, the aforementioned methods have largely adopted techniques from NeRF extension research that were designed for natural scene tasks or they developed methods for volumetric reconstruction in a more general sense.
However, there is significant potential for specialized techniques to enhance the accuracy and efficiency of CBCT image reconstruction. 
Specifically, we argue that the soft and hard tissue parts are inherently different, both radiologically and physiologically, disentangling and imposing regularization on these components will be beneficial to the learning process.

In this work, we propose a novel approach from a distinctly new perspective to address the challenges in sparse-view CBCT reconstruction by disentangling the reconstruction to hard and soft tissue and imposing disentangled supervision, resulting in improved quality and faster convergence. 

\section{Methods}
In this section, we first review the fundamentals of the CBCT reconstruction problem and the neural rendering techniques. We then introduce the detailed descriptions of the actual algorithm, 
accompanied by comprehensive explanations of the underlying intuition that drives our approach.

\subsection{Problem Formulation}
The image formation part of a CBCT machine mainly consists of an X-ray source and a plane-shaped detector.
During image capturing, the two components rotate around the patient simultaneously. 
X-ray source emits many cone-shaped X-ray beams, and the detector captures the remaining X-ray signals after being absorbed by the patient from multiple angles~\cite{scarfe2008whatiscbct}.
Those captured X-ray projections are used to reconstruct the volumetric image.

For each ray $\mathbf{r}(t) \in \mathbb{R}^3$, emitted from the X-ray tube, parameterized by $t$.
The received X-ray intensity $I$ from the detector is submitted to the Lambert-Beers law:
\begin{equation}
    I = I_0e^{-\int_{t\in l}\sigma(\mathbf{r}(t))dt},
    \label{eq:cbct_formation}
\end{equation}
where $I_0$ represents the intensity originated from the X-ray source, $l$ denotes the range of $t$ that defines the ray-path, and $\sigma$ represents the intensity field of the volume under consideration. 
The primary objective of CBCT reconstruction is to resolve the field $\sigma$ (\textit{i.e.,} attenuation coefficient). If the intensity field is to be represented as a neural field, then the reconstruction problem becomes to train this neural network based on the observations.

One who is familiar with NeRF~\cite{mildenhall2021nerf} will surely realize that these two have similar formulations.
In NeRF, the image formation process is expressed as:
\begin{align}
     \mathbf{C}(\mathbf{r}) &= 
        \int_{t_n}^{t_f}\mathbf{T}(t)
        \sigma(\mathbf{r}(t))
        \mathbf{c}(\mathbf{r}(t), \mathbf{d})dt \label{eq:nerf_formation},\\
    \text{where }
    \mathbf{T}(t) &= e^{-\int_{t_n}^{t}\sigma(\mathbf{r}(t))dt}.
    \label{eq:nerf_formation_transmittance}
\end{align}
If light source emit a mono-color $I_0$ at the farthest point and disregard any other emissions along the ray path ($\boldsymbol{c}(\boldsymbol{r}(t), \boldsymbol{d}) = 0$, if $t \neq t_f$), 
equation \eqref{eq:nerf_formation} simplifies to solely the transmittance component $\mathbf{T}$ computed from that farthest point. Consequently, this simplification makes it equivalent to \eqref{eq:cbct_formation}. 

In practice, the CT reconstruction is commonly solved in logarithmic space, 
\textit{i.e.,} to teat the projection as integration along the ray:
\begin{equation}
    \ln(I_0) - \ln(I) = \int_{t_n}^{t_f}\sigma(\mathbf{r}(t))dt.
\end{equation}
Thus, the rendering process can be treated as simply a summation of the sampled points along the ray.

\begin{figure*}[ht!]
    \centering
    \includegraphics[width=.95\linewidth, trim=0 0 0 5,clip]{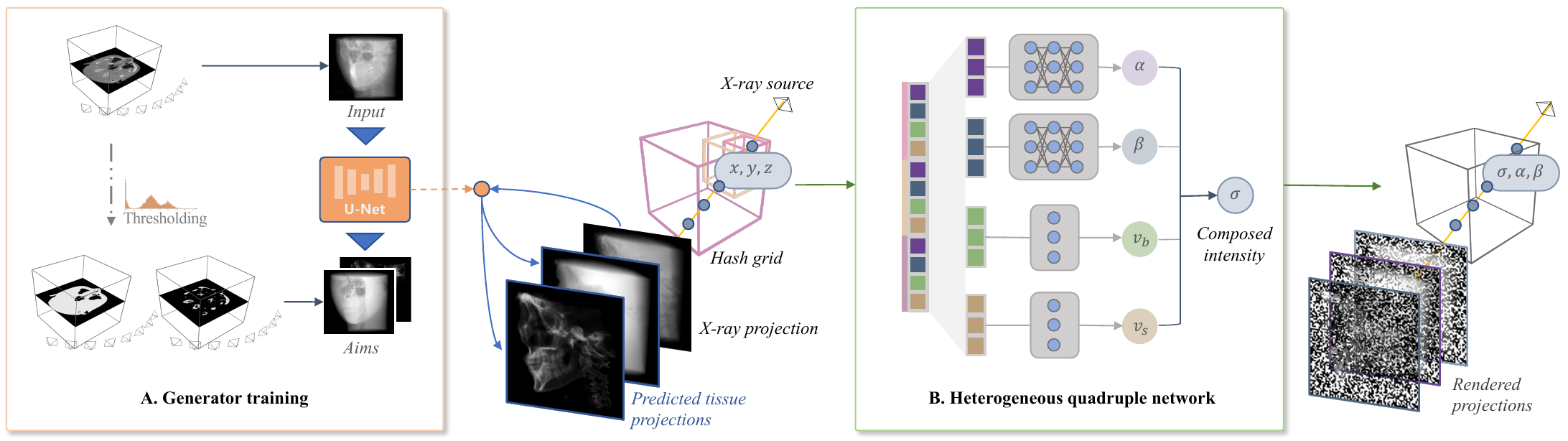}
    % \vspace{-1mm}
    \caption{\small
    Overview of the training scheme. 
    (A) The tissue projections are synthesized from segmentation obtained from thresholding, which is used to train the generator to predict the tissue projections out of X-ray projection. 
    (B) The heterogeneous quadruple network is designed. We split the network into four separate branches, in order to encourage the network to learn diverse representations. The capacity of the \textit{value} branches is limited by a smaller number of parameters, resulting in smooth textures.
    }
    \label{fig:workflow}
\end{figure*}

\subsection{Intensity Field Decomposition} \label{sec:decomposition}
We start off by re-formulating the attenuation field $\sigma$ into disentangled shapes and textures, which is a key feature of our method. 
We express a tissue as a multiplicative mask with the texture, 
so that the mask defines the spacial boundary of the tissue, segregating the high-frequency boundary and low-frequency texture; 
Specifically, for any 3D coordinate $\boldsymbol{x} \in \mathbb{R}^3$ we express the intensity value $\sigma(\boldsymbol{x})$ as follows:
\begin{equation}
    \sigma(\boldsymbol{x}) = (\alpha(\boldsymbol{x}) + \epsilon)
    (\beta(\boldsymbol{x}) v_b(\boldsymbol{x}) + v_s(\boldsymbol{x})),
    \label{eq:decomposition}
\end{equation}
where $\alpha, \beta, v_b, v_s: \mathbb{R}^3 \mapsto \mathbb{R}$.
$\alpha$ represent the mask of the object (analogous to \textit{$\alpha$-matting}), 
$\beta$ is the mask of the hard tissue (\textit{i.e.,} bone), 
$v_b$ and $v_s$ are textures (values) of the hard and soft tissue, respectively. 
The learned scalar $\epsilon$ is a small, position-irrelevant number introduced to serve as a ground base, 
so that low-intensity values out of the $\alpha$ mask, such as air, can be well defined. 
This decomposition allows us to split the intensity field into 4 components: $\alpha$, $\beta$, $v_b$, and $v_s$ (As shown in Fig~\ref{fig:teaser}). 

There are several advantages of this disentangled representation, which is also the motivation behind the decomposition:
1. CBCT has a low intensity variation on soft tissue (\textit{i.e.,} lower soft tissue contrast)~\cite{pauwels2015technicalofcbct}, indicating that the texture of soft tissue should be relatively smooth. 
By the disentanglement, we can explicitly constrain the $v_b$ and $v_s$ to have more low-frequency components, 
while keeping the shape $\alpha$ and $\beta$ intact.
2. We use sigmoid activation in the last layers of the network so that the four components are in the range of $(0,1)$. Ideally, the shape components $\alpha$ and $\beta$ are binary segmentation masks of the tissue. Compared to directly predicting $\sigma$, it is easier for the network to learn binary outputs with the sigmoid activation, leading to enhanced convergence. 
3. Provided that the network can produce a smooth field for $v_b$ and $v_s$, 
this representation guarantees that the bony tissue will exhibit a higher intensity value than the soft tissue locally (\textit{i.e.,} $\sigma_{\alpha\rightarrow1, \beta\rightarrow1} > \sigma_{\alpha\rightarrow1, \beta\rightarrow0}$). 
As the hard tissue parts can be seen as an addictive texture onto the soft tissue texture, and the \eqref{eq:decomposition} will degenerate into only the soft tissue components where $\beta$ is zero. 
By leveraging these insights, our method decouples the reconstruction into shape and texture components, enables better control over the reconstruction process, and leads to more interpretable results. 

In addition, we predict the tissue mask projection as an additional supervision for improved reconstruction.  
The idea is based on the fact that X-rays produce a clear image of the hard tissue and thus the bony structures can be evaluated out of X-ray projections. 
% A single X-ray image is also a routine clinical diagnostic tool in orthodontics, where a cephalometry is used to evaluate the shape of the skull~\cite{devereux2011cepImportant}. 
The details of the tissue supervision will be introduced in the following section.

\subsection{Tissue Supervision}
\subsubsection{Obtaining Volume Segmentation Masks}
CBCT images typically exhibit distinct intensity distributions between soft and hard tissue.  
 %~\cite{halazonetis2005threedcep, flugge2017registration}. 
For this reason, when the volume image is known, it is straightforward to obtain the tissue regions by selecting appropriate thresholds. 

However, the volume image is the ultimate goal of the reconstruction and is unknown. 
As discussed above, we propose estimating the accumulated tissue mask projections based on real X-ray projection, which can then be used as additional supervision during training.

\subsubsection{Predicting Segmentation Mask Projections}
Let $\Sigma$ be the accumulated intensity $\sigma$ on the projection panel of size $M \times N$, $\mathcal{\Alpha}$ and $\mathcal{\Beta}$ be the accumulated $\alpha$ and $\beta$ projections, with $\Sigma, \mathcal{\Alpha}, \mathcal{\Beta} \in \mathbb{R}^{M\times N}$. 
We use a U-Net generator $\mathcal{G}$ to predict estimated tissue mask projections $\mathcal{\Alpha}_{pred}$ and $\mathcal{\Beta}_{pred}$ based on real X-ray projection $\Sigma_{gt}$, 
so that 
\begin{equation}
    \mathcal{G}(\Sigma_{gt}) = 
    [\mathcal{\Alpha}_{pred}, \mathcal{\Beta}_{pred}]. 
\end{equation}
The generator can be trained with CBCT volume images where the initial segmentation mask is obtained using thresholding. 
A forward projection process as described in \eqref{eq:cbct_formation} was employed to create the dataset (Fig.~\ref{fig:workflow}~A). 

\begin{table*}[ht!]\Huge
    \caption{Quantitative Results on the oral-maxillofacial dataset. 
    We show the difference ($\Delta$) compared to SART~\cite{andersen1984sart} in bracket, and highlight the increases of at least 0.5 PSNR or 0.1 SSIM in blue, while decreases are in red.}
    \resizebox{\linewidth}{!}{\begin{tabular}{l|cccccc}
% \toprule
\hline
   &\multicolumn{1}{c}{\textbf{10 View}}
   &\multicolumn{1}{c}{\textbf{15 View}}
   &\multicolumn{1}{c}{\textbf{20 View}}
   &\multicolumn{1}{c}{\textbf{30 View}}
   &\multicolumn{1}{c}{\textbf{40 View}}
   &\multicolumn{1}{c}{\textbf{60 View}}
   \\
Model&PSNR$\uparrow${\Large\ ($\Delta$)}/SSIM$\uparrow${\Large\ ($\Delta$)}&PSNR$\uparrow${\Large\ ($\Delta$)}/SSIM$\uparrow${\Large\ ($\Delta$)}&PSNR$\uparrow${\Large\ ($\Delta$)}/SSIM$\uparrow${\Large\ ($\Delta$)}&PSNR$\uparrow${\Large\ ($\Delta$)}/SSIM$\uparrow${\Large\ ($\Delta$)}&PSNR$\uparrow${\Large\ ($\Delta$)}/SSIM$\uparrow${\Large\ ($\Delta$)}&PSNR$\uparrow${\Large\ ($\Delta$)}/SSIM$\uparrow${\Large\ ($\Delta$)}
\\
  % \midrule
  \hline
SART\cite{andersen1984sart}&21.3\ \ \ \ \ \ /0.56\ \ \ \ \ \ \ \,&21.5\ \ \ \ \ \ /0.58\ \ \ \ \ \ \ \,&21.8\ \ \ \ \ \ \,/0.61\ \ \ \ \ \ \ \,&23.9\ \ \ \ \ \ \,/0.66\ \ \ \ \ \ \ \,&25.3\ \ \ \ \ \ \,/0.69\ \ \ \ \ \ \ \,&27.4\ \ \ \ \ \ \,/0.76\ \ \ \ \ \ \ \,
\\FDK\cite{feldkamp1984fdk}&16.0\textcolor{red}{{\Large\ (-5.3)}\ }/0.12\textcolor{red}{{\Large\ (-0.44)}}&17.8\textcolor{red}{{\Large\ (-3.7)}\ }/0.16\textcolor{red}{{\Large\ (-0.42)}}&19.2\textcolor{red}{{\Large\ (-2.6)}\ }/0.21\textcolor{red}{{\Large\ (-0.40)}}&21.3\textcolor{red}{{\Large\ (-2.6)}\ }/0.29\textcolor{red}{{\Large\ (-0.37)}}&22.9\textcolor{red}{{\Large\ (-2.4)}\,}/0.37\textcolor{red}{{\Large\ (-0.32)}}&25.0\textcolor{red}{{\Large\ (-2.4)}\,}/0.49\textcolor{red}{{\Large\ (-0.27)}}
\\PatRecon\cite{Shen2019PatRecon}&22.1\textcolor{NavyBlue}{{\Large\ (+0.8)}}/0.62{\Large\ (+0.06)}&21.8{\Large\ (+0.3)}/0.61{\Large\ (+0.03)}&22.3\textcolor{NavyBlue}{{\Large\ (+0.5)}}/0.63{\Large\ (+0.02)}&22.1\textcolor{red}{{\Large\ (-1.8)}\ }/0.62{\Large\ (-0.04)}&21.4\textcolor{red}{{\Large\ (-3.9)}\,}/0.60{\Large\ (-0.09)}&21.9\textcolor{red}{{\Large\ (-5.5)}\,}/0.61\textcolor{red}{{\Large\ (-0.15)}}
\\NeAT\cite{ruckert2022neat}&21.7{\Large\ (+0.4)}/0.65{\Large\ (+0.09)}&22.8\textcolor{NavyBlue}{{\Large\ (+1.3)}}/0.69\textcolor{NavyBlue}{{\Large\ (+0.11)}}&23.5\textcolor{NavyBlue}{{\Large\ (+1.7)}}/0.72\textcolor{NavyBlue}{{\Large\ (+0.11)}}&24.3{\Large\ (+0.4)}/0.77\textcolor{NavyBlue}{{\Large\ (+0.11)}}&24.7\textcolor{red}{{\Large\ (-0.6)}}/0.80\textcolor{NavyBlue}{{\Large\ (+0.11)}}&25.1\textcolor{red}{{\Large\ (-2.3)}}/0.84{\Large\ (+0.08)}
\\DIF-Net\cite{lin2023difnet}&25.8\textcolor{NavyBlue}{{\Large\ (+4.5)}}/0.70\textcolor{NavyBlue}{{\Large\ (+0.14)}}&27.2\textcolor{NavyBlue}{{\Large\ (+5.7)}}/0.74\textcolor{NavyBlue}{{\Large\ (+0.16)}\ }&27.7\textcolor{NavyBlue}{{\Large\ (+5.9)}}/0.73\textcolor{NavyBlue}{{\Large\ (+0.12)}}&28.7\textcolor{NavyBlue}{{\Large\ (+4.8)}}/0.75{\Large\ (+0.09)}&29.3\textcolor{NavyBlue}{{\Large\ (+4.0)}}/0.77{\Large\ (+0.08)}&30.1\textcolor{NavyBlue}{{\Large\ (+2.7)}}/0.78{\Large\ (+0.02)}
\\NAF\cite{zha2022naf}&25.9\textcolor{NavyBlue}{{\Large\ (+4.6)}}/0.72\textcolor{NavyBlue}{{\Large\ (+0.16)}}&28.1\textcolor{NavyBlue}{{\Large\ (+6.6)}}/0.77\textcolor{NavyBlue}{{\Large\ (+0.19)}}&29.5\textcolor{NavyBlue}{{\Large\ (+7.7)}}/0.81\textcolor{NavyBlue}{{\Large\ (+0.20)}}&31.5\textcolor{NavyBlue}{{\Large\ (+7.6)}}/0.86\textcolor{NavyBlue}{{\Large\ (+0.20)}}&32.8\textcolor{NavyBlue}{{\Large\ (+7.5)}}/0.89\textcolor{NavyBlue}{{\Large\ (+0.20)}}&34.2\textcolor{NavyBlue}{{\Large\ (+6.8)}}/0.92\textcolor{NavyBlue}{{\Large\ (+0.16)}}
\\MLP(hash)&26.2\textcolor{NavyBlue}{{\Large\ (+4.9)}}/0.75\textcolor{NavyBlue}{{\Large\ (+0.19)}}&28.2\textcolor{NavyBlue}{{\Large\ (+6.7)}}/0.79\textcolor{NavyBlue}{{\Large\ (+0.21)}}&29.7\textcolor{NavyBlue}{{\Large\ (+7.9)}}/0.82\textcolor{NavyBlue}{{\Large\ (+0.21)}}&31.8\textcolor{NavyBlue}{{\Large\ (+7.9)}}/0.86\textcolor{NavyBlue}{{\Large\ (+0.20)}}&33.5\textcolor{NavyBlue}{{\Large\ (+8.2)}}/0.89\textcolor{NavyBlue}{{\Large\ (+0.20)}}&\cellcolor{BestBg}{36.2}\textcolor{NavyBlue}{{\Large\ (+8.8)}}/\cellcolor{SecBg}{0.93}\textcolor{NavyBlue}{{\Large\ (+0.17)}}
\\\textbf{TNT(const.$\lambda$)}&\cellcolor{BestBg}{\textbf{27.8}}\textcolor{NavyBlue}{\textbf{{\Large\ (+6.5)}}}/\cellcolor{BestBg}{\textbf{0.81}}\textcolor{NavyBlue}{\textbf{{\Large\ (+0.25)}}}&\cellcolor{SecBg}{29.7}\textcolor{NavyBlue}{{\Large\ (+8.2)}}/\cellcolor{SecBg}{0.84}\textcolor{NavyBlue}{{\Large\ (+0.26)}}&\cellcolor{SecBg}{31.0}\textcolor{NavyBlue}{{\Large\ (+9.2)}}/\cellcolor{SecBg}{0.86}\textcolor{NavyBlue}{{\Large\ (+0.25)}}&\cellcolor{SecBg}{32.7}\textcolor{NavyBlue}{{\Large\ (+8.8)}}/\cellcolor{SecBg}{0.88}\textcolor{NavyBlue}{{\Large\ (+0.22)}}&\cellcolor{SecBg}{34.2}\textcolor{NavyBlue}{{\Large\ (+8.9)}}/\cellcolor{SecBg}{0.90}\textcolor{NavyBlue}{{\Large\ (+0.21)}}&36.0\textcolor{NavyBlue}{{\Large\ (+8.6)}}/0.93\textcolor{NavyBlue}{{\Large\ (+0.17)}}
\\\textbf{TNT}&\cellcolor{SecBg}{27.7}\textcolor{NavyBlue}{{\Large\ (+6.4)}}/\cellcolor{SecBg}{0.81}\textcolor{NavyBlue}{{\Large\ (+0.25)}}&\cellcolor{BestBg}{\textbf{29.9}}\textcolor{NavyBlue}{\textbf{{\Large\ (+8.4)}}}/\cellcolor{BestBg}{\textbf{0.85}}\textcolor{NavyBlue}{\textbf{{\Large\ (+0.27)}}}&\cellcolor{BestBg}{\textbf{31.5}}\textcolor{NavyBlue}{\textbf{{\Large\ (+9.7)}}}/\cellcolor{BestBg}{\textbf{0.87}}\textcolor{NavyBlue}{\textbf{{\Large\ (+0.26)}}}&\cellcolor{BestBg}{\textbf{33.7}}\textcolor{NavyBlue}{\textbf{{\Large\ (+9.8)}}}/\cellcolor{BestBg}{\textbf{0.90}}\textcolor{NavyBlue}{\textbf{{\Large\ (+0.24)}}}&\cellcolor{BestBg}{\textbf{35.3}}\textcolor{NavyBlue}{\textbf{{\Large\ (+10.0)}}}/\cellcolor{BestBg}{\textbf{0.92}}\textcolor{NavyBlue}{\textbf{{\Large\ (+0.23)}}}&\cellcolor{BestBg}{\textbf{37.3}}\textcolor{NavyBlue}{\textbf{{\Large\ (+9.9)}}}/\cellcolor{BestBg}{\textbf{0.94}}\textcolor{NavyBlue}{\textbf{{\Large\ (+0.18)}}}
\\
% \bottomrule
\hline
\end{tabular}}
    \label{tab:metrics}
\end{table*}

\subsubsection{Loss}
Since the estimated tissue mask projections may not be accurate and lack multi-view consistency, the inclusion of a strong tissue supervision can potentially lead to false results. 
To address this issue, we propose the following loss function: 
\begin{equation}
    \mathcal{L} = \mathcal{L}_1(\hat{\Sigma}, \Sigma_{gt}) 
        + \lambda(t) \mathcal{L}_2[(\hat{\mathcal{\Alpha}}, \hat{\mathcal{\Beta}}), (\mathcal{\Alpha}_{pred}, \mathcal{\Beta}_{pred})],
    \label{eq:loss}
\end{equation}
where $\hat{\Sigma}, \hat{\mathcal{\Alpha}}, \hat{\mathcal{\Beta}}$ are accumulated outputs from the network. 
The \eqref{eq:loss} is designed with the following two considerations:

First, We use the mean absolute error loss $\mathcal{L}_1$ on the X-ray projection $\Sigma$ and mean square loss $\mathcal{L}_2$ on the tissue projections $\mathcal{\Alpha}$ and $\mathcal{\Beta}$.
The loss will penalize for discrepancies between the accumulated intensity projections $\hat{\Sigma}$ and the ground truth X-ray projection $\Sigma_{gt}$ 
to ensure the predicted tissue mask aligns closely with the ground truth. 
To the other end, we incorporate mean square error loss between the accumulated tissue mask $\alpha$ and $\beta$ projections ($\hat{\mathcal{\Alpha}}$, $\hat{\mathcal{\Beta}}$) and their corresponding predicted values ($\mathcal{\Alpha}_{pred}$, $\mathcal{\Beta}_{pred}$), imposing a strong penalty on outliers while keeping the penalty relatively weak for closer matches. 

Second, we add a time-dependent regulation factor $\lambda$ to the tissue supervision:
\begin{equation}
    \lambda(t) = \max(0, 1-\frac{t}{k T})^2\lambda_0, 
\end{equation}
where $t$ and $T$ are the current training iteration and total iterations, respectively.
$\lambda_0$ is a constant. 
% Empirically, we set $\tau$ to 50\% of the total training iterations, 
Empirically, we set $\lambda_0=5$ and $k=0.5$, in this way, the training will be guided by the possibly inaccurate tissue supervision at the start to force the network to learn a desired pattern of the tissue masks ($\alpha$ and $\beta$). 
In the later stage of training, we let the network accommodate to the optimal distribution by itself, without tissue supervision. 

By doing so, we compensate for the inaccuracy and lack of multi-view consistency in the predicted projections and encourage the network to self-supervise and predict a multi-view consistent tissue mask that is close to the ground truth.

\subsection{Network Structure}
Based on the aforementioned disentangled representation, it is evident that a quadruple network output is necessary. 
However, simply adapting the last layer of the MLP to a 4-channel output and imposing tissue-based supervision is not enough to produce optimal results. 
The reason is two-fold: 
First, the shape and value components are inherently different in nature (\textit{e.g.,} in terms of value range and frequency component), making it challenging for a fully connected MLP to learn such a diverse representation since all features are always shared between layers, especially when strongly parameterized embedding is adopted. 
% It's especially true when strongly parameterized embedding is adopted, the high-frequency details are mostly learned within the embedding, which makes it even harder for the network to learn semantically diverse representations. 
Second, the shape branches ($\alpha, \beta$) of the network are supervised by the predicted tissue mask projections, but there are no restrictions on the value branches ($v_b, v_s$), which makes it easy for the network to generate predictions that contradict the assumption of low-frequency outcomes on those branches. 

To achieve the above-mentioned objectives, we propose to utilize a quadruple network that is composed of 4 separated embeddings and small MLPs, to learn explicitly disentangled features. 
In addition, instead of manually designing regularization, we introduce smaller networks on those branches, which limits the complexity of the networks, also reduces the number of parameters for accelerated training (refer to Fig.~\ref{fig:workflow}~B). 

As previously discussed, grid-based representations are popular in neural implicit scene representation due to their ability in reducing network size, accelerate network training, and yield promising results, which have already been adopted in related tasks~\cite{zha2022naf, fang2022snaf, ruckert2022neat}. 
Following previous works, we choose the hash grid \cite{muller2022instantngp} as our positional encoding. 

Since the network produces shapes and values associated with soft and hard tissue, and the training is guided by tissue supervision. 
We refer to the network, together with its training strategy, as {\em Tissue-guided Neural Tomography (TNT)}.

\section{Implementation Details}

\noindent\textbf{Network structure: } 
We employ TransUNet with ResNet-50 backbone~\cite{chen2021transunet} as generator to predict tissue projections. 
In practice, we implement the encoding as splitting the outcome of a single hash grid with 4-dimension feature per level (refer to Fig.~\ref{fig:workflow}). This approach is essentially equivalent to using 4 distinct hash grid encodings with 1-dimension feature per level. 
For the value branches ($v_b, v_s$), we use a linear mapping of the embedding output followed by an activation to minimize the network capacity, while preserving the training efficiency. 

\noindent\textbf{Training setup: }
We train our reconstruction network with 10,000 iterations per scene, which leverages all available views within each iteration.
For each projection, we randomly select a batch size of 1,024 pixels for rendering. Along each ray, we uniformly sample 576 points.
An Adam optimizer with a learning rate of 3e-4 is used to train the network.

\section{Experiments}

\begin{table*}[ht!]\Huge
    \caption{
    Quantitative Results on the LIDC-IDRI Dataset. 
    We show the difference ($\Delta$) compared to SART in bracket, and highlight the increases of at least 0.5 PSNR or 0.05 SSIM in blue, while decreases are in red.
    }
    \resizebox{\linewidth}{!}{\begin{tabular}{l|cccccc}
% \toprule
\hline
&\multicolumn{1}{c}{\textbf{10 View}}
&\multicolumn{1}{c}{\textbf{15 View}}
&\multicolumn{1}{c}{\textbf{20 View}}
&\multicolumn{1}{c}{\textbf{30 View}}
&\multicolumn{1}{c}{\textbf{40 View}}
&\multicolumn{1}{c}{\textbf{60 View}}
\\
Model
&PSNR$\uparrow${\Large\ ($\Delta$)}/SSIM$\uparrow${\Large\ ($\Delta$)}
&PSNR$\uparrow${\Large\ ($\Delta$)}/SSIM$\uparrow${\Large\ ($\Delta$)}
&PSNR$\uparrow${\Large\ ($\Delta$)}/SSIM$\uparrow${\Large\ ($\Delta$)}
&PSNR$\uparrow${\Large\ ($\Delta$)}/SSIM$\uparrow${\Large\ ($\Delta$)}
&PSNR$\uparrow${\Large\ ($\Delta$)}/SSIM$\uparrow${\Large\ ($\Delta$)}
&PSNR$\uparrow${\Large\ ($\Delta$)}/SSIM$\uparrow${\Large\ ($\Delta$)}
\\ 
% \midrule
\hline
SART~\cite{andersen1984sart}
	% &25.29\ \ \ \ /0.58\ \ \ \ \ \ \ 
	% &26.79\ \ \ \ /0.63\ \ \ \ \ \ \ 
	% &27.67\ \ \ \ /0.66\ \ \ \ \ \ \ 
	% &29.07\ \ \ \ /0.71\ \ \ \ \ \ \ 
	% &30.04\ \ \ \ /0.75\ \ \ \ \ \ \ 
	% &31.53\ \ \ \ /0.81\ \ \ \ \ \ \ 
    &25.3\ \ \ \ \ \ \,/0.58\ \ \ \ \ \ \ 
	&26.8\ \ \ \ \ \ \,/0.63\ \ \ \ \ \ \ 
	&27.7\ \ \ \ \ \ /0.66\ \ \ \ \ \ \ 
	&29.0\ \ \ \ \ \ /0.71\ \ \ \ \ \ \ 
	&30.0\ \ \ \ \ \ /0.75\ \ \ \ \ \ \ 
	&31.5\ \ \ \ \ \ /0.81\ \ \ \ \ \ \ 
\\FDK~\cite{feldkamp1984fdk}
	&17.2\textcolor{red}{\Large\ (-8.1)}\,/0.16\textcolor{red}{\Large\ (-0.42)}
	&19.4\textcolor{red}{\Large\ (-7.4)}/0.22\textcolor{red}{\Large\ (-0.41)}
	&20.9\textcolor{red}{\Large\ (-6.8)}/0.27\textcolor{red}{\Large\ (-0.39)}
	&22.8\textcolor{red}{\Large\ (-6.3)}/0.34\textcolor{red}{\Large\ (-0.37)}
	&24.0\textcolor{red}{\Large\ (-6.0)}/0.40\textcolor{red}{\Large\ (-0.35)}
	&25.5\textcolor{red}{\Large\ (-6.1)}/0.49\textcolor{red}{\Large\ (-0.32)}
\\PatRecon~\cite{Shen2019PatRecon}
	&25.0\textcolor{black}{\Large\ (-0.3)}/0.60\textcolor{black}{\Large\ (+0.02)}
	&25.0\textcolor{red}{\Large\ (-1.8)}/0.60\textcolor{black}{\Large\ (-0.03)}
	&25.0\textcolor{red}{\Large\ (-2.6)}/0.60\textcolor{red}{\Large\ (-0.06)}
	&25.1\textcolor{red}{\Large\ (-4.0)}/0.60\textcolor{red}{\Large\ (-0.11)}
	&25.0\textcolor{red}{\Large\ (-5.0)}/0.61\textcolor{red}{\Large\ (-0.14)}
	&24.8\textcolor{red}{\Large\ (-6.7)}/0.60\textcolor{red}{\Large\ (-0.21)}
\\NeAT~\cite{ruckert2022neat}
	&23.1\textcolor{red}{\Large\ (-2.2)}/0.58\textcolor{black}{\Large\ (+0.00)}
	&23.9\textcolor{red}{\Large\ (-2.9)}/0.62\textcolor{black}{\Large\ (-0.01)}
	&22.6\textcolor{red}{\Large\ (-5.1)}/0.60\textcolor{red}{\Large\ (-0.06)}
	&24.7\textcolor{red}{\Large\ (-4.3)}/0.67\textcolor{black}{\Large\ (-0.04)}
	&24.9\textcolor{red}{\Large\ (-5.1)}/0.69\textcolor{red}{\Large\ (-0.06)}
	&25.0\textcolor{red}{\Large\ (-6.5)}/0.71\textcolor{red}{\Large\ (-0.10)}
\\DIF-Net~\cite{lin2023difnet}
	&25.1\textcolor{black}{\Large\ (-0.2)}/0.56\textcolor{black}{\Large\ (-0.02)}
	&28.9\textcolor{NavyBlue}{\Large\ (+2.1)}/0.69\textcolor{NavyBlue}{\Large\ (+0.06)}
	&29.3\textcolor{NavyBlue}{\Large\ (+1.7)}/0.72\textcolor{NavyBlue}{\Large\ (+0.06)}
	&30.0\textcolor{NavyBlue}{\Large\ (+0.9)}/0.72\textcolor{black}{\Large\ (+0.01)}
	&30.4\textcolor{black}{\Large\ (+0.4)}/0.74\textcolor{black}{\Large\ (-0.01)}
	&31.2\textcolor{black}{\Large\ (-0.4)}/0.76\textcolor{red}{\Large\ (-0.05)}
\\NAF~\cite{zha2022naf}
	&28.1\textcolor{NavyBlue}{\Large\ (+2.8)}/0.66\textcolor{NavyBlue}{\Large\ (+0.08)}
	&29.5\textcolor{NavyBlue}{\Large\ (+2.7)}/0.70\textcolor{NavyBlue}{\Large\ (+0.07)}
	&30.2\textcolor{NavyBlue}{\Large\ (+2.6)}/0.72\textcolor{NavyBlue}{\Large\ (+0.06)}
	&31.3\textcolor{NavyBlue}{\Large\ (+2.2)}/0.76\textcolor{NavyBlue}{\Large\ (+0.05)}
	&32.0\textcolor{NavyBlue}{\Large\ (+2.0)}/0.78\textcolor{black}{\Large\ (+0.03)}
	&\cellcolor{SecBg}{33.0\textcolor{NavyBlue}{\Large\ (+1.4)}/0.81\textcolor{black}{\Large\ (+0.00)}}
\\MLP (hash)
	&\cellcolor{SecBg}{27.5\textcolor{NavyBlue}{\Large\ (+2.2)}/0.66\textcolor{NavyBlue}{\Large\ (+0.08)}}
	&\cellcolor{SecBg}{29.6\textcolor{NavyBlue}{\Large\ (+2.8)}/0.71\textcolor{NavyBlue}{\Large\ (+0.08)}}
	&\cellcolor{SecBg}{30.4\textcolor{NavyBlue}{\Large\ (+2.8)}/0.73\textcolor{NavyBlue}{\Large\ (+0.07)}}
	&\cellcolor{SecBg}{31.9\textcolor{NavyBlue}{\Large\ (+2.8)}/0.77\textcolor{NavyBlue}{\Large\ (+0.06)}}
	&\cellcolor{SecBg}{32.8\textcolor{NavyBlue}{\Large\ (+2.8)}/0.80\textcolor{NavyBlue}{\Large\ (+0.05)}}
	&\cellcolor{BestBg}{\textbf{34.4}\textcolor{NavyBlue}{\textbf{\Large\ (+2.9)}}/\textbf{0.84}\textcolor{black}{\textbf{\Large\ (+0.03)}}} 
\\\textbf{TNT}
	&\cellcolor{BestBg}{\textbf{29.0}\textcolor{NavyBlue}{\textbf{\Large\ (+3.7)}}/\textbf{0.72}\textcolor{NavyBlue}{\textbf{\Large\ (+0.14)}}}
	&\cellcolor{BestBg}{\textbf{30.8}\textcolor{NavyBlue}{\textbf{\Large\ (+4.0)}}/\textbf{0.76}\textcolor{NavyBlue}{\textbf{\Large\ (+0.13)}}}
	&\cellcolor{BestBg}{\textbf{31.6}\textcolor{NavyBlue}{\textbf{\Large\ (+4.0)}}/\textbf{0.78}\textcolor{NavyBlue}{\textbf{\Large\ (+0.12)}}}
	&\cellcolor{BestBg}{\textbf{32.8}\textcolor{NavyBlue}{\textbf{\Large\ (+3.7)}}/\textbf{0.80}\textcolor{NavyBlue}{\textbf{\Large\ (+0.09)}}}
	&\cellcolor{BestBg}{\textbf{33.5}\textcolor{NavyBlue}{\textbf{\Large\ (+3.5)}}/\textbf{0.82}\textcolor{NavyBlue}{\textbf{\Large\ (+0.07)}}}
	&\cellcolor{BestBg}{\textbf{34.4}\textcolor{NavyBlue}{\textbf{\Large\ (+2.9)}}/\textbf{0.84}\textcolor{black}{\textbf{\Large\ (+0.03)}}}
\\ 
% \bottomrule
\hline
\end{tabular}}
    \label{tab:metrics_lidc}
\end{table*}

\subsection{Experiment Settings}

\noindent\textbf{Dataset: } 
We evaluated our method on two datasets: 
1. \textit{Oral-maxillofacial dataset}: 
    We collected 150 maxillofacial CBCT volumes due to its significant clinical relevance to CBCT imaging. 
2. \textit{The LIDC-IDRI dataset}~\cite{armato2011lidc}: 
    We also tested the proposed method on the LIDC-IDRI dataset, 209 subjects were excluded from the total 1018 CT subjects, leaving 809 volumes for the experiments\footnote{
    This curation was due to the inherent variability in the number of slices present in the dataset, which is designed for Multi-slice CT imaging. 
    Given that our experiment is tailored for Cone-beam CT (CBCT) we excluded subjects with fewer than 128 slices to maintain the desired sharpness of the volume. 
    }.

All volumetric images were resized into 3D cube volume with dimensions of \(256 \times 256 \times 256\) for the experiments. For each dataset, we randomly reserve 10 volumes to evaluate the reconstruction quality, while the others were used for training the U-Net generator, as well as the counterpart baseline models (when pre-training is required). 

\noindent\textbf{Data Simulation: }
To simulate the CBCT projection process, we create a circular trajectory on the central coronal plane of the volumes and generate 120 projections with dimensions of \(512 \times 512 \) pixels in 180 degrees. The reconstruction network was evaluated using uniformly sampled projections (number of projections ranging from 10 - 60), and the tissue prediction network was trained on all 120 projections.

\noindent\textbf{Metrics: } The peak signal-to-noise ratio (PSNR) and structural similarity index measure (SSIM) were chosen to evaluate the quality of reconstruction, and higher values of PSNR and SSIM indicate better quality of reconstruction. 

\noindent\textbf{Baselines: }we compare our method with the following baseline methods:
(1) Two classical methods: 
\textit{FDK}\cite{feldkamp1984fdk} and 
\textit{SART}~\cite{andersen1984sart}; % implemented by TIGRE toolbox~\cite{Biguri_2016tigre}; 
(2) An end-to-end CNN-based sparse-view CT reconstruction method, 
\textit{PatRecon}\cite{Shen2019PatRecon}; 
(3) Four coordinate-based neural intensity field models for CBCT imaging: 
\textit{NeAT}~\cite{ruckert2022neat}, \textit{NAF}~\cite{zha2022naf}, \textit{DIF-Net}~\cite{lin2023difnet}, and \textit{MLP (hash)}. 
Where MLP (hash) is a model using the same architecture as NAF, and uses a larger MLP and a more parameter-rich hash grid, enhancing model capacity. 
For a fair comparison, the model parameters of MLP (hash) are matched with those of TNT, ensuring similar representational power across methods.

\noindent\textbf{Ethics statement: }
The study protocol was approved by the institutional ethics committee of the Hospital of Stomatology, Wuhan University (No. 2023[B46]). 
The study adhered to the ethical considerations outlined by the committee, including participant privacy, and data protection measures.

\subsection{Comparison to Baselines}

We extensively evaluated our method and compared it with various baselines under different levels of viewing sparsities.

% ================================================================

\begin{figure}[ht]
    \centering
    \includegraphics[width=.95\linewidth, trim=0 0 0 24,clip]{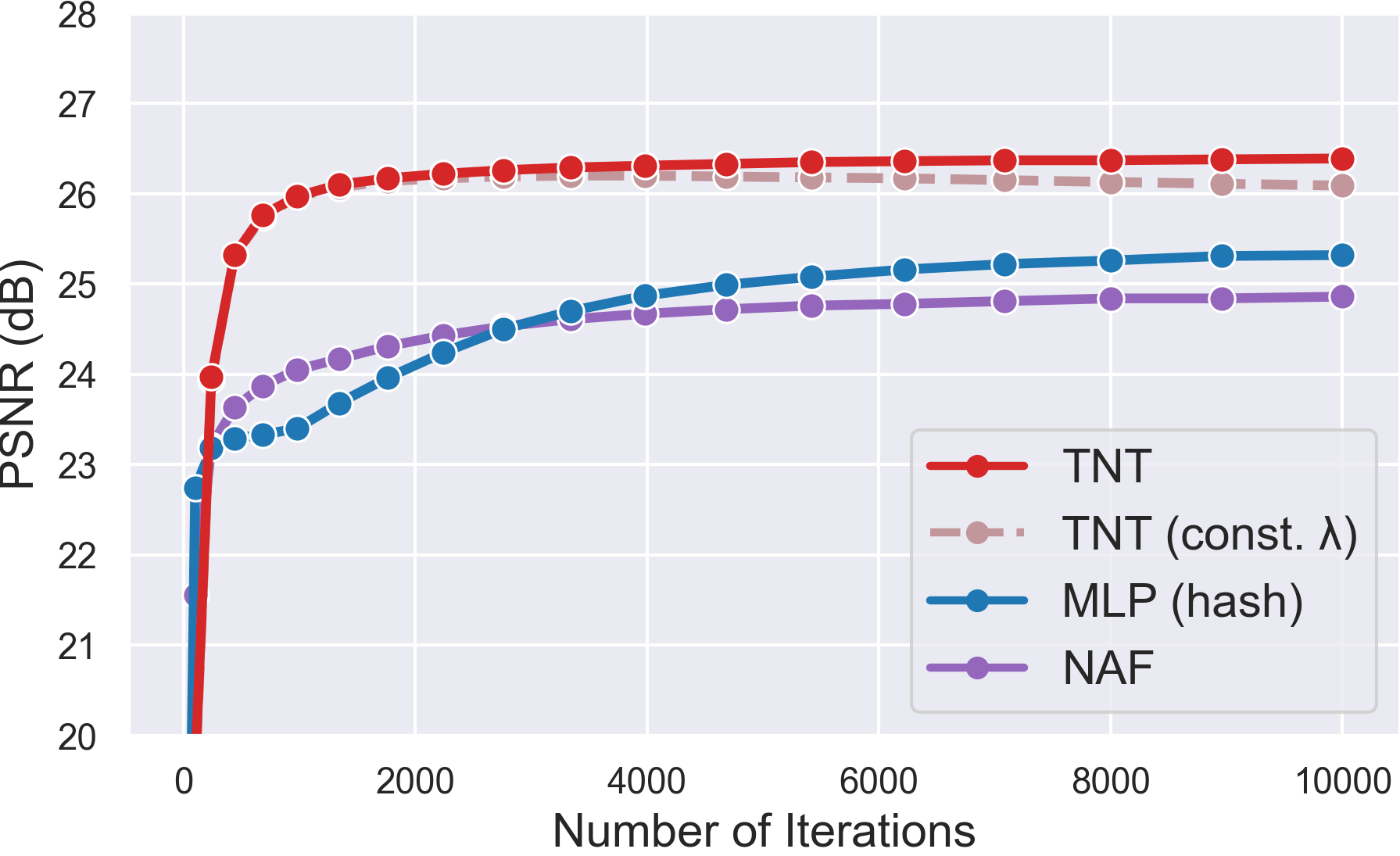}
    \caption{\small
    PSNR performance across iterations. 
    }
    \label{fig:epoch2psnr}
\end{figure}

\begin{table}[b]
    \caption{Reconstruction Within 30-minute Period}
    \resizebox{\linewidth}{!}{\begin{tabular}{lcccccc} 
% \toprule
\hline
 & \multicolumn{1}{c}{\textbf{15-view}} & \multicolumn{1}{c}{\textbf{30-view}}  & \multicolumn{1}{c}{\textbf{40-view}}  \\
Model  &  PSNR $\uparrow$/SSIM $\uparrow$ &  PSNR $\uparrow$/SSIM $\uparrow$  &  PSNR $\uparrow$/SSIM $\uparrow$ \\
% \midrule
\hline
NeAT \cite{ruckert2022neat}
&22.67/0.68	
&24.07/0.74
&24.45/0.78
\\
NAF \cite{zha2022naf}
&28.04/0.77
&31.25/0.85
&32.50/0.89
\\
MLP (hash)
&27.55/0.76
&30.40/0.82
&31.99/0.86
\\
\textbf{TNT}
&\textbf{29.79}/\textbf{0.85}
&\textbf{32.74}/\textbf{0.89}
&\textbf{33.86}/\textbf{0.91}
\\
% \bottomrule
\hline
\end{tabular}}
    \label{tab:tab_time}
\end{table}

\subsubsection{Quantitative Results.}

A detailed reconstruction metrics from the two datasets are shown in Table~\ref{tab:metrics} and Table~\ref{tab:metrics_lidc}. The PSNR/SSIM values are averaged across 10 test volumes of different individuals. 

As shown in the Tables, the methods relying on coordinate-based neural field outperform traditional ones and CNN-based end to end method. Notably, our method exhibits superior performance than all the comparable methods in sparse-view conditions on both datasets (10 - 60 views).

\subsubsection{Qualitative Results.}

\begin{figure*}[hbt!]
    \centering
    \includegraphics[width=.98\linewidth, trim=0 0 0 820,clip]{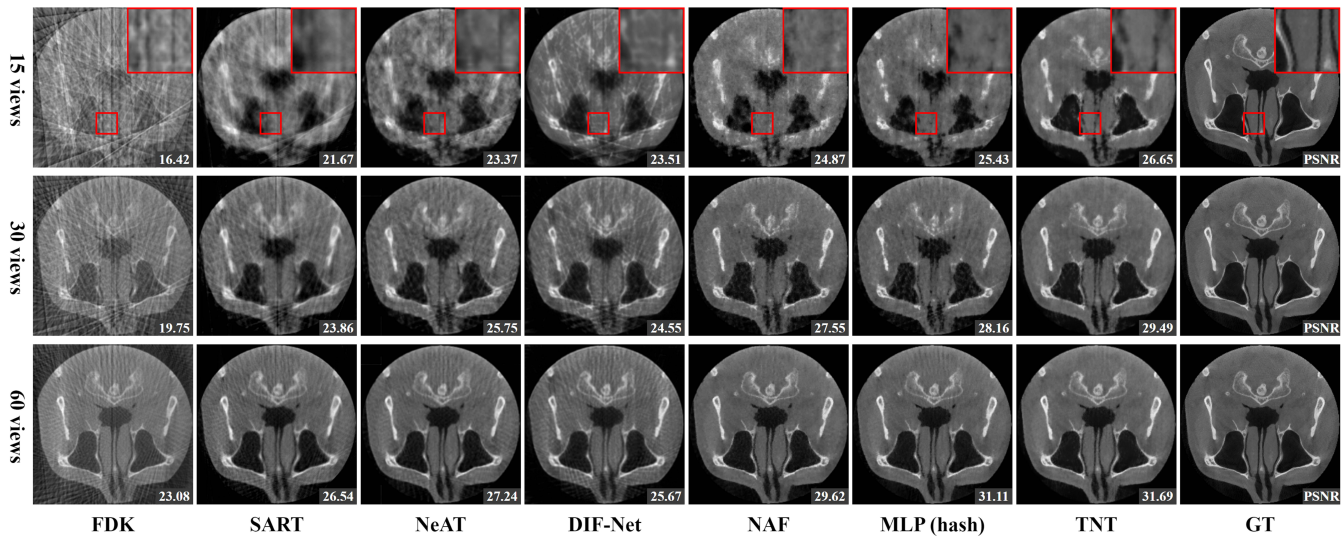}
    \vspace{-1mm}
    \caption{\small
    Qualitative results on the oral-maxillofacial dataset.
    }
    \label{fig:result3}
\end{figure*}

\begin{figure*}[hbt!]
    \centering
    \includegraphics[width=.98\linewidth, trim=0 0 0 210,clip]{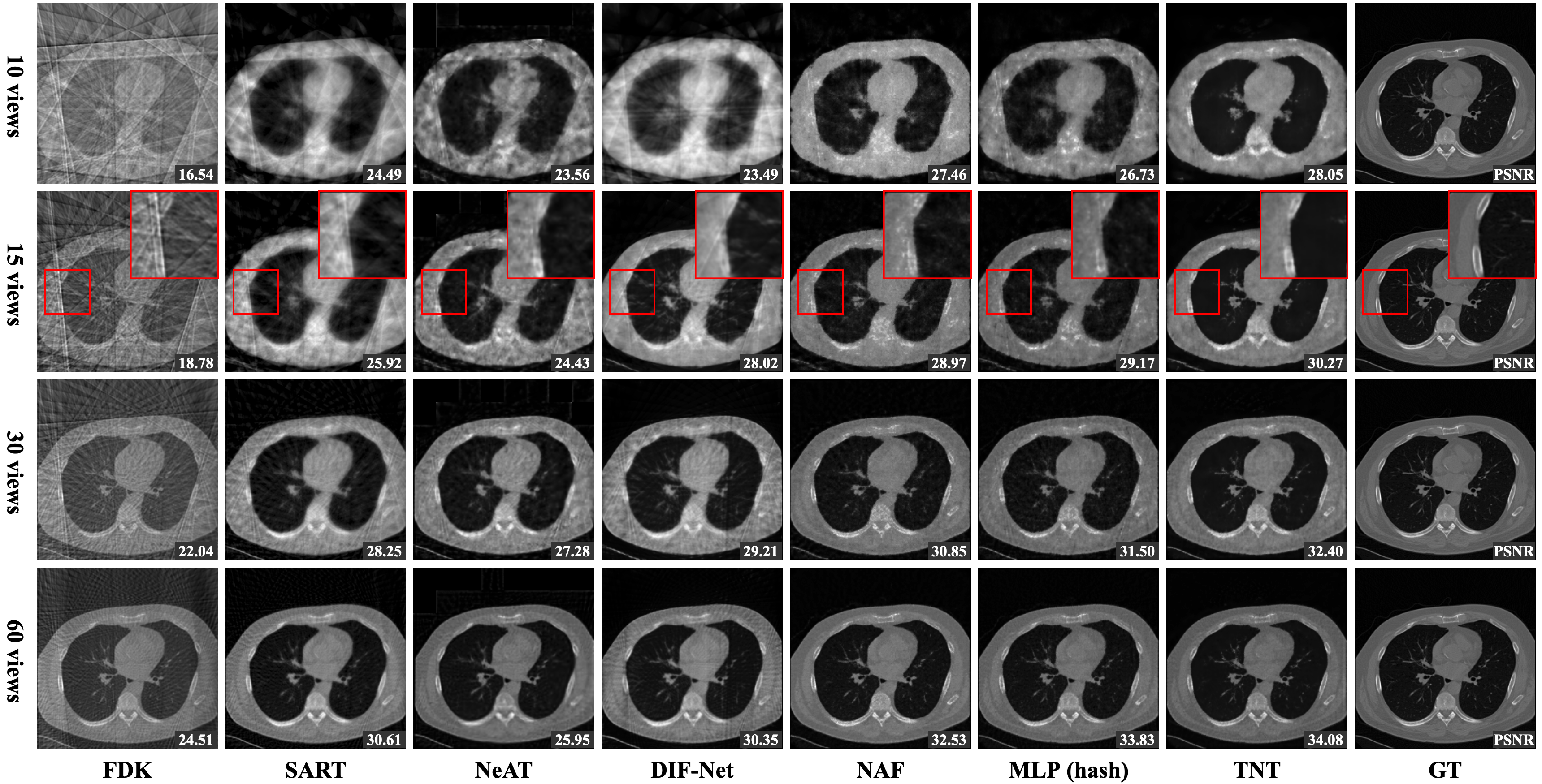}
    \vspace{-1mm}
    \caption{\small
    Qualitative results on the LIDC-IDRI dataset. 
    }
    \label{fig:lidc-sectional}
\end{figure*}

\begin{figure}
    \centering
    \includegraphics[width=\linewidth]{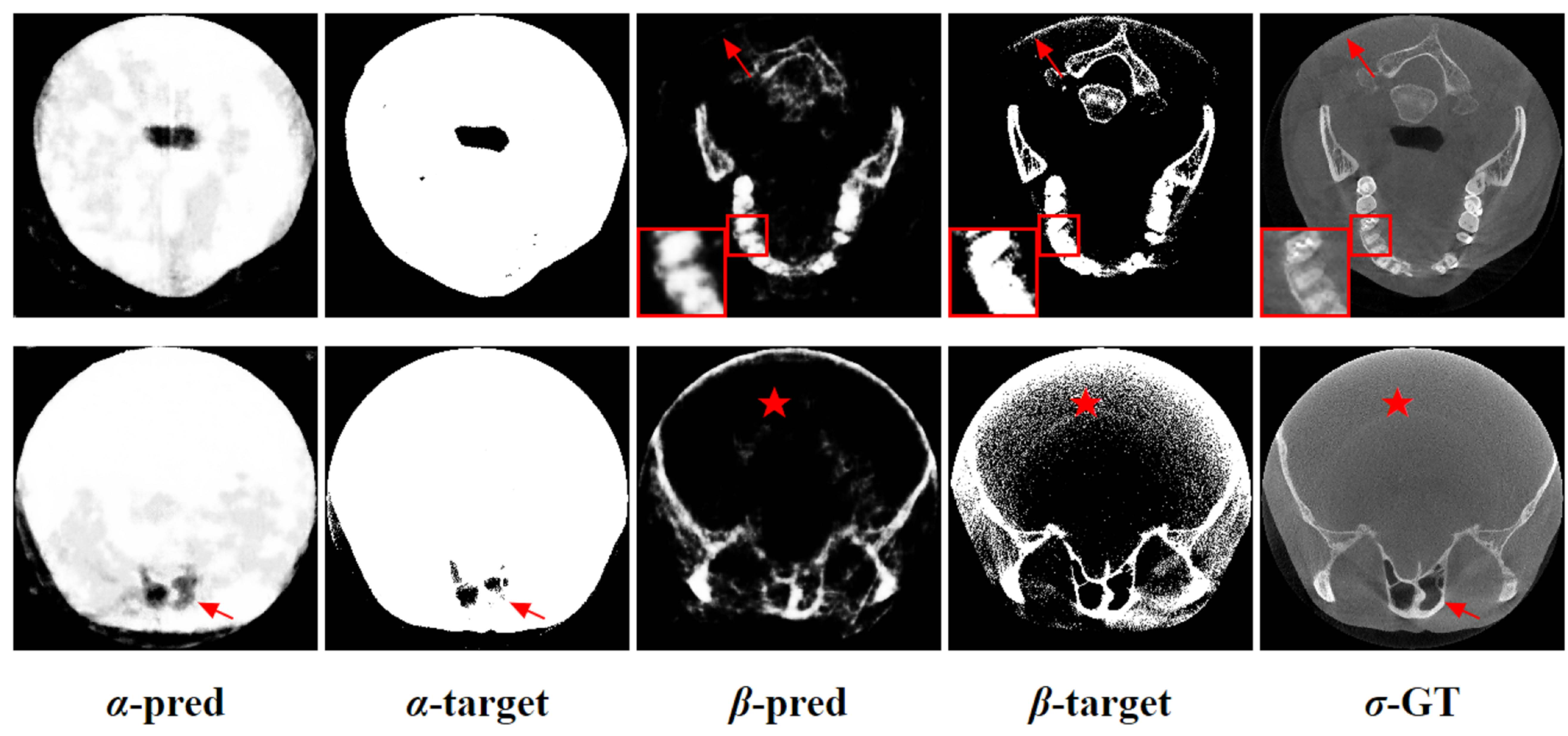}
    \caption{\small
    Learned tissue fields. The predicted tissue masks are closer to the real tissue regions such as teeth and sphenoid sinus than the target threshold segmentation.
    }
    \label{fig:result_threshold82}
\end{figure}

Cross-sectional views of reconstructed volumes on two datasets are displayed in Fig.~\ref{fig:result3} and Fig.~\ref{fig:lidc-sectional}. 
Among the methods, FDK struggles to generate distinguishable images on sparse view inputs. 
SART, NeAT, DIF-Net, and NAF can generate correct shapes but introduce significant noise and artifacts when the number of projections is limited. 
MLP with a larger number of parameters shows improved reconstruction quality, which however still cannot produce a decent outcome with less than 30 projections. 
In comparison, the proposed TNT demonstrates far better image quality under sparse views. 
As illustrated in the figure, the TNT produces clearer and sharper anatomical structures such as nasal concha and meatus on only 15 views, and achieves satisfactory reconstruction in 30 views. 

A further investigation of the underlying tissue representations learned from 20 projections is shown in Fig.~\ref{fig:result_threshold82}. 
The model learned semantically meaningful fields for both soft and hard tissue masks. 
Notably, the learned tissue segmentation made several corrections to the targets,  revealing clearer teeth and sphenoid sinus shape (refer to markers on Fig.~\ref{fig:result_threshold82}).

\subsubsection{Training Efficiency}

The reconstructed volume PSNRs during training is illustrated in Fig.~\ref{fig:epoch2psnr}. 
One can observe that the proposed TNT converges way faster than its counterpart models by achieving the highest reconstruction quality with only about 1/10 of the training iterations. 
To further assess the training efficiency of different neural rendering models, we conducted a comparative analysis by considering the reconstruction quality achieved within a fixed training time.
As presented in Table.~\ref{tab:tab_time}, TNT outperforms the competitive methods within a 30-minute training period.

\subsection{Experiments on Out-of-distribution Data}
\begin{figure}
    \centering
    \includegraphics[width=\linewidth]{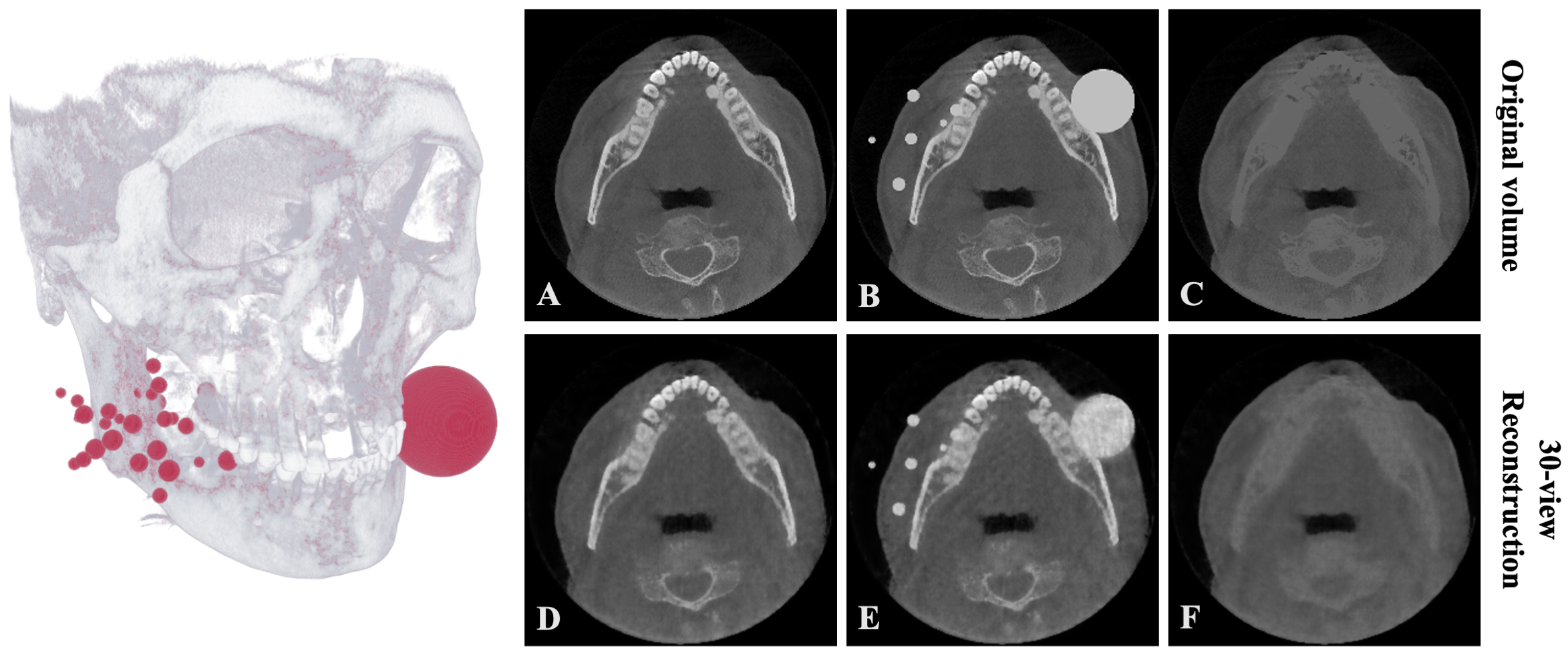}
    \vspace{-3mm}
    \caption{\small
    Out-of-distribution data.
    The aims of the experiment include:
    (A) The original CBCT; 
    (B) Foreign bodies added to the cheek region;
    (C) Intensity clipped to remove hard tissue.
    The second row presents the reconstruction of TNT under 30 views. 
    }
    \label{fig:ood-sectional}
\end{figure}

We established two cases to examine the resilience of the proposed method against out-of-distribution data: 
In the first case, we introduced ball-shaped objects to the cheek regions (as shown on the left side of Fig.~\ref{fig:ood-sectional}).  
The second case is a head CBCT with only soft tissue, 
where the hard tissue was removed by clipping the intensity. 
Both of these are extreme cases that were not covered by the training data of the tissue-prediction network, 
nor will they appear in clinical practice.

\begin{table}
    \centering
    \caption{Evaluation on the Out-of-distribution Data.}
    \resizebox{\linewidth}{!}{
\begin{tabular}{|c|c|c|c|c|}
    \hline

    & \multicolumn{2}{c|}{\textbf{With objects}} 
    & \multicolumn{2}{c|}{\textbf{Soft only}} \\

    \cline{2-5}

    & \textbf{TNT} & \textbf{MLP(hash)} 
    & \textbf{TNT} & \textbf{MLP(hash)} \\

    \hline

    \textbf{10 views} 
        & \textbf{24.21}/\textbf{0.66} & 23.22/0.61  
        & 28.75/0.75 & \textbf{29.48}/\textbf{0.76} \\
    \textbf{15 views} 
        & \textbf{26.13}/\textbf{0.70} & 24.85/0.65  
        & 30.99/0.79 & \textbf{31.15}/0.79 \\
    \textbf{20 views} 
        & \textbf{27.49}/\textbf{0.73} & 26.27/0.69 
        & 32.15/\textbf{0.81} & \textbf{32.19}/0.80 \\
    \textbf{30 views} 
        & \textbf{29.05}/\textbf{0.77} & 27.74/0.72 
        & \textbf{33.43}/0.83 & 33.36/0.83 \\
    \textbf{40 views} 
        & \textbf{30.10}/\textbf{0.79} & 28.86/0.75 
        & \textbf{34.32}/0.84 & 34.06/0.84 \\

    \hline

\end{tabular}}
    \label{tab:exp_ood}
\end{table}

Fig. \ref{fig:ood-sectional} and Table \ref{tab:exp_ood} demonstrate the results. 
In the case of artificial objects, 
the reconstruction exhibits slightly increased noise 
(Fig.~\ref{fig:ood-sectional}~(E)). 
However, the result presents clear boundaries of the subjects. 
Table~\ref{tab:exp_ood} further reveals that our method is still superior under this condition. 
In the case involving only soft tissue, 
while the metric decreases (Table~\ref{tab:exp_ood}), the result remains on par with the comparative method. 
Moreover, no tissue-related artifacts, such as the undesired presence of hard tissue, are presented (Fig.~\ref{fig:ood-sectional}~(F)).

\subsection{Ablation Studies}

We further investigated several designing factors that may influence the model performance. 

\subsubsection{Effectiveness of the Annealing Supervision}

In the later stage of training, the regulation factor $\lambda$ was reduced to let the network evolve without tissue supervision. 
We argue that inaccurate tissue estimation may hinder the reconstruction, for the network has a tendency to overfit the tissue projection and ultimately lead to artifacts. The self-supervision is critical in coping with this inaccuracy. 
To validate this insight, we trained TNT with a constant $\lambda$ (\textit{i.e., }$\lambda \defeq \lambda_0$), we call this network TNT (const. $\lambda$). The qualitative results of this training scheme is reported in Table~\ref{tab:metrics}.
Even with constant tissue regulation, TNT (const. $\lambda$) achieves good reconstruction in most cases (Table~\ref{tab:metrics}). 
Only when the network has access to sufficient information about the object (over 60 views) is TNT (const. $\lambda$) outperformed by other methods. This outcome is reasonable, as estimated tissue supervision can provide additional information, but it may have an adverse effect by forcing the network to fit view-inconsistent tissue projections. 

Fig.~\ref{fig:epoch2psnr} also shows this adverse effect with decreased PSNR in the latter stage of training. 
Luckily, the proposed annealing tissue supervision mitigates this limitation, leading to improved reconstruction quality in almost all situations.

\begin{figure}
    \centering
    \includegraphics[width=\linewidth]{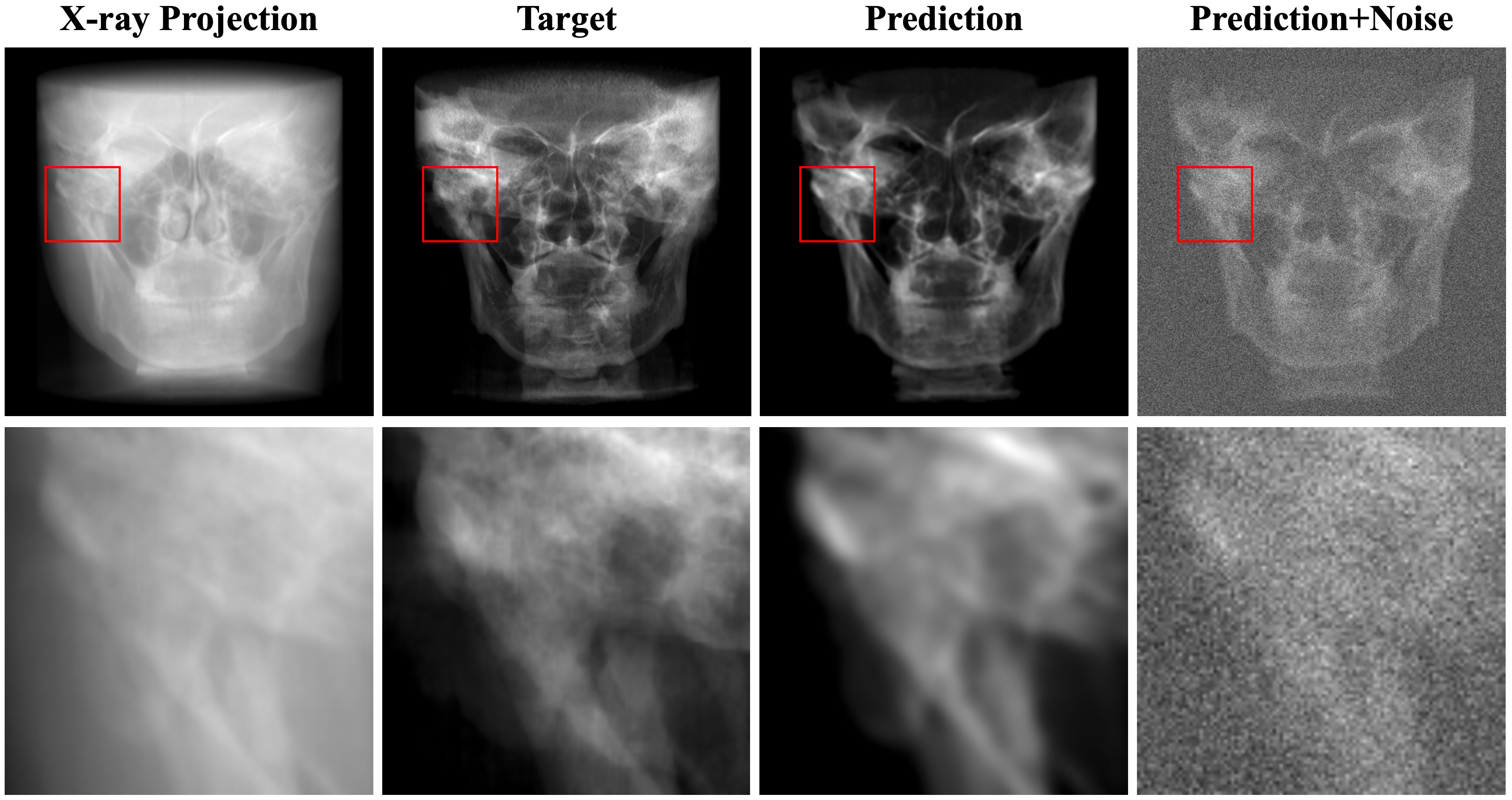}
    \vspace{-3mm}
    \caption{\small
    Three types of tissue projections in for ablation: 
    1) {\em target}: the projections of the threshold-ed ground truth volume image; 
    2) {\em prediction}: the output of the tissue generator network; 
    3) {\em prediction+noise}: the noise-perturbed predicted tissue projections.
    }
    \label{fig:supervision_schemes}
\end{figure}

\subsubsection{Impact of Tissue Supervision Accuracy}
\begin{figure}[ht!]
    \centering
    \includegraphics[width=\linewidth]{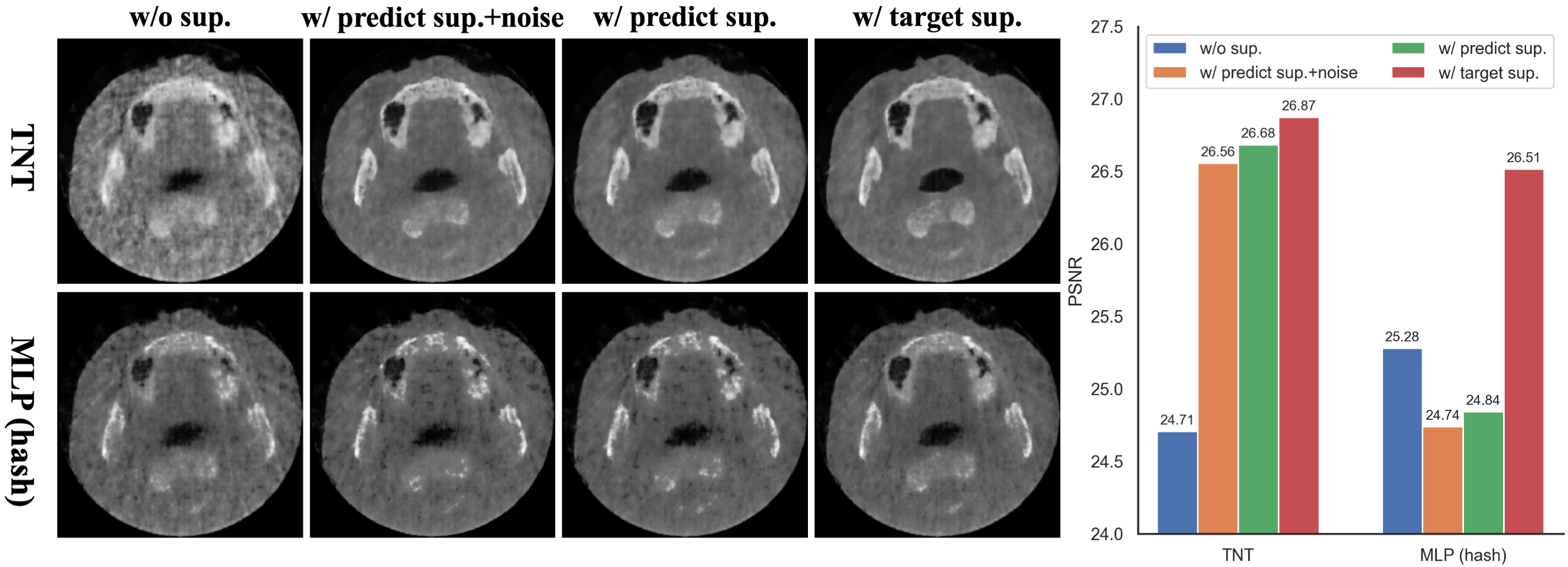}
    \caption{\small
    Results from 15 views using different supervision schemes:  
    no tissue supervision (w/o sup., blue), 
    with noised predicted tissue supervision (w/ predict sup.+noise, orange), 
    with predicted tissue supervision (w/ predict sup., green), 
    and with target tissue supervision (w/ target sup., red). 
    }
    \label{fig:ablation_sup_slice}
\end{figure}

It is clear that the predicted tissue projection is a suboptimal substitute for the unknown tissue projections. 
Whether obtaining view consistent supervision would enhance image quality remains an unresolved question. 
Moreover, the impact of introducing inaccurate tissue prediction is another question that requires further investigation.
To address these questions, instead of using the prediction from the generator, we attempted to guide the reconstruction with the target tissue projections directly (obtained by rendering threshold-ed volume image, Fig.~\ref{fig:workflow}~A). 
Additionally, we experimented with guiding the reconstruction using tissue predictions perturbed with noise. 
An illustration of projections from different supervision schemes is provided in Fig.~\ref{fig:supervision_schemes}. 
The results of the experiments are presented in Fig.~\ref{fig:ablation_sup_slice}. 
As expected, TNT achieves better reconstruction quality under the target viewing consistent supervision. 
However, the improvement is not substantial compared to the predicted tissue supervision. 
Furthermore, the reconstruction quality remains high even in the presence of noisy predictions. 
The network's resilience to variations from tissue prediction should be attributed to the inherent smoothness of the neural field, as well as the loss design, and the annealing tissue supervision scheme. 
In practice, it is impossible to obtain the target projections. 
Consequently, the experiment shows that generated tissue projections can be considered a suitable alternative.

\subsubsection{Tissue Supervision on Network with Single Output}
Since the target tissue segmentation are obtained by an unary operation (thresholding), we can easily apply it to the network output and obtain the tissue projections, which can then be subjected to supervision. 

Specifically, by setting desired thresholds, denoted as $T_\alpha, T_\beta$, in accordance with the generator training scheme, we can obtain the rendered tissue projections with:

\begin{equation}
   Acc(i) = \int_{t_n}^{t_f}\tau_i (\sigma(t))dt, 
\end{equation}
where $Acc(i)$ is the accumulated projections (\textit{i.e.,} $\hat{\mathcal{\Alpha}}, \hat{\mathcal{\Beta}}$ in \eqref{eq:loss}). $\tau_i$ is the thresholding process, with $i\in \{\alpha, \beta\}$:
\begin{equation}
    \tau_i(\sigma) =
  \begin{cases}
    1 \text{, if } \sigma \geq T_i\vspace{3pt}, \\
    0 \text{, otherwise.}
  \end{cases}
\end{equation}

With the accumulated projections, a network with a single output can be trained with tissue supervision. 
Following this approach, we trained MLP (hash) with tissue supervision and compared it with TNT, visualizations of the outcome are also illustrated in Fig.~\ref{fig:ablation_sup_slice}.
The reconstruction quality of the network with a single output is improved, provided that target tissue projections can be obtained. 
However, despite having the same number of parameters, the MLP achieved inferior results compared to TNT. 
Furthermore, the MLP's performance is impaired by the predicted tissue supervision, 
The results indicate that the TNT is more resilient to inaccurate tissue predictions and is better suited to tissue supervision.

\subsubsection{Disentangled Output without Tissue Supervision}
As regulatory factor $\lambda$ was set to zero at the halfway point of training, one may naturally question whether the disentanglement alone can lead to improved reconstruction quality. In other words, the loss function becomes solely the data fidelity term:
\begin{equation}
    \mathcal{L} = \mathcal{L}_1(\hat{\Sigma}, \Sigma_{gt}). 
\end{equation}
Again, we compared TNT without tissue supervision (w/o sup.) to that of MLP (hash), the results are shown in Fig.~\ref{fig:ablation_sup_slice} as blue. 
TNT faces difficulties in generating high-quality reconstructions in the absence of tissue supervision. We propose that this issue stems from the inherent challenge of enabling the four network heads to collaboratively learn the intended pattern for distinguishing between tissues. 
Moreover, the multiplicative relationship among these heads~\eqref{eq:decomposition} further exacerbates the complexity of this learning process. 
When tissue supervision is presented, it serves as guidance and enable the initial patterns to be learned efficiently. 

\section{Limitations}
One limitation of the proposed method is that it only applies regularization to known views.  
This less intrusive prior intervention will be insufficient under extremely sparse observations. 
Nevertheless, the diagnostic demands in medical imaging suggest that extremely sparse view CT reconstruction may be less feasible,
and our method can produce decent results when given moderately sparse views.

% \section{Conclusions}
% The reconstruction of CBCT under sparse-view projections is a valuable yet challenging task. 
% In this work, we are inspired by a basic clinical observation that bony tissue can be clearly distinguished from X-ray projections, we thus tackle the challenging reconstruction problem with a decomposed neural intensity field. 
% We presented TNT, a heterogeneous quadruple network with a tissue-guided training strategy. Our approach significantly improves the reconstruction quality and accelerates the convergence process of the neural rendering based CBCT reconstruction, without any prior assumption on unknown views. 
% We also conducted a detailed ablation analysis of our proposed TNT, which indicates that the promising performance is attributed to the novel model design, disentangled supervision, and time-dependent training schemes. 
% We believe this work will provide valuable insight for the related neural rendering based medical imaging tasks.

\normalsize
\bibliography{citations}

\begin{thebibliography}{10}
\providecommand{\url}[1]{#1}
\csname url@samestyle\endcsname
\providecommand{\newblock}{\relax}
\providecommand{\bibinfo}[2]{#2}
\providecommand{\BIBentrySTDinterwordspacing}{\spaceskip=0pt\relax}
\providecommand{\BIBentryALTinterwordstretchfactor}{4}
\providecommand{\BIBentryALTinterwordspacing}{\spaceskip=\fontdimen2\font plus
\BIBentryALTinterwordstretchfactor\fontdimen3\font minus \fontdimen4\font\relax}
\providecommand{\BIBforeignlanguage}[2]{{%
\expandafter\ifx\csname l@#1\endcsname\relax
\typeout{** WARNING: IEEEtran.bst: No hyphenation pattern has been}%
\typeout{** loaded for the language `#1'. Using the pattern for}%
\typeout{** the default language instead.}%
\else
\language=\csname l@#1\endcsname
\fi
#2}}
\providecommand{\BIBdecl}{\relax}
\BIBdecl

\bibitem{kaasalainen2021cbctreview}
T.~Kaasalainen \emph{et~al.}, ``Dental cone beam ct: An updated review,'' \emph{Physica Medica}, vol.~88, pp. 193--217, 2021.

\bibitem{stratis2019cbctconcern}
A.~Stratis \emph{et~al.}, ``The growing concern of radiation dose in paediatric dental and maxillofacial cbct: an easy guide for daily practice,'' \emph{European Radiology}, vol.~29, pp. 7009--7018, 2019.

\bibitem{feldkamp1984fdk}
L.~A. Feldkamp \emph{et~al.}, ``Practical cone-beam algorithm,'' \emph{Journal of The Optical Society of America A-optics Image Science and Vision}, vol.~1, pp. 612--619, 1984.

\bibitem{andersen1984sart}
A.~H. Andersen and A.~C. Kak, ``Simultaneous algebraic reconstruction technique (sart): a superior implementation of the art algorithm,'' \emph{Ultrasonic imaging}, vol.~6, no.~1, pp. 81--94, 1984.

\bibitem{arridge2019solving}
S.~Arridge \emph{et~al.}, ``Solving inverse problems using data-driven models,'' \emph{Acta Numerica}, vol.~28, pp. 1--174, 2019.

\bibitem{zhang2018ctregreview}
H.~Zhang \emph{et~al.}, ``Regularization strategies in statistical image reconstruction of low-dose x-ray ct: A review,'' \emph{Medical Physics}, vol.~45, no.~10, pp. e886--e907, 2018.

\bibitem{zhu2018imrec_nature}
B.~Zhu \emph{et~al.}, ``Image reconstruction by domain-transform manifold learning,'' \emph{Nature}, vol. 555, no. 7697, pp. 487--492, Mar 2018.

\bibitem{genzel2023dlmedrecrobust}
M.~Genzel \emph{et~al.}, ``Solving inverse problems with deep neural networks – robustness included?'' \emph{IEEE Transactions on Pattern Analysis and Machine Intelligence}, vol.~45, no.~1, pp. 1119--1134, Jan 2023.

\bibitem{wang2020dl}
G.~Wang \emph{et~al.}, ``Deep learning for tomographic image reconstruction,'' \emph{Nature Machine Intelligence}, vol.~2, pp. 737--748, 12 2020.

\bibitem{liu2023deepeit}
D.~Liu \emph{et~al.}, ``Deepeit: Deep image prior enabled electrical impedance tomography,'' \emph{IEEE Transactions on Pattern Analysis and Machine Intelligence}, vol.~45, no.~08, pp. 9627--9638, aug 2023.

\bibitem{tewari2022neuralrenderreview}
A.~Tewari \emph{et~al.}, ``Advances in neural rendering,'' \emph{Computer Graphics Forum}, vol.~41, no.~2, pp. 703--735, 2022.

\bibitem{ruckert2022neat}
D.~R\"{u}ckert \emph{et~al.}, ``Neat: Neural adaptive tomography,'' \emph{ACM Transactions on Graphics}, vol.~41, no.~4, jul 2022.

\bibitem{zha2022naf}
R.~Zha \emph{et~al.}, ``Naf: Neural attenuation fields for sparse-view cbct reconstruction,'' in \emph{International Conference on Medical Image Computing and Computer-Assisted Intervention}, 2022, pp. 442--452.

\bibitem{fang2022snaf}
Y.~Fang \emph{et~al.}, ``Snaf: Sparse-view cbct reconstruction with neural attenuation fields,'' \emph{arXiv:2211.17048}, 2022.

\bibitem{abril2022mednerf}
A.~Corona-Figueroa \emph{et~al.}, ``Mednerf: Medical neural radiance fields for reconstructing 3d-aware ct-projections from a single x-ray,'' in \emph{International Conference of the IEEE Engineering in Medicine \& Biology Society}.\hskip 1em plus 0.5em minus 0.4em\relax IEEE, 2022, pp. 3843--3848.

\bibitem{lin2023difnet}
Y.~Lin \emph{et~al.}, ``Learning deep intensity field for extremely sparse-view cbct reconstruction,'' in \emph{Medical Image Computing and Computer Assisted Intervention -- MICCAI 2023}, H.~Greenspan \emph{et~al.}, Eds.\hskip 1em plus 0.5em minus 0.4em\relax Cham: Springer Nature Switzerland, 2023, pp. 13--23.

\bibitem{sun2023xrayct}
Y.~Sun \emph{et~al.}, ``Ct reconstruction from few planar x-rays with application towards low-resource radiotherapy,'' in \emph{Deep Generative Models}, A.~Mukhopadhyay \emph{et~al.}, Eds.\hskip 1em plus 0.5em minus 0.4em\relax Cham: Springer Nature Switzerland, 2024, pp. 225--234.

\bibitem{liu2024svcbct}
Z.~Liu \emph{et~al.}, ``Geometry-aware attenuation learning for sparse-view cbct reconstruction,'' \emph{IEEE Transactions on Medical Imaging}, 2024.

\bibitem{yu2022monosdf}
Z.~Yu \emph{et~al.}, ``Monosdf: Exploring monocular geometric cues for neural implicit surface reconstruction,'' \emph{Advances in neural information processing systems}, vol.~35, pp. 25\,018--25\,032, 2022.

\bibitem{wei2023nerfingmvs}
Y.~Wei \emph{et~al.}, ``Depth-guided optimization of neural radiance fields for indoor multi-view stereo,'' \emph{IEEE Transactions on Pattern Analysis and Machine Intelligence}, vol.~45, no.~9, pp. 10\,835--10\,849, Sep. 2023.

\bibitem{wang2023sparsenerf}
G.~Wang \emph{et~al.}, ``Sparsenerf: Distilling depth ranking for few-shot novel view synthesis,'' in \emph{IEEE/CVF International Conference on Computer Vision (ICCV)}, 2023.

\bibitem{somraj2023vipnerf}
N.~Somraj and R.~Soundararajan, ``Vip-nerf: Visibility prior for sparse input neural radiance fields,'' in \emph{ACM SIGGRAPH 2023 Conference Proceedings}, ser. SIGGRAPH '23, 2023.

\bibitem{Michael2022RegNeRF}
M.~Niemeyer \emph{et~al.}, ``Regnerf: Regularizing neural radiance fields for view synthesis from sparse inputs,'' in \emph{Proceedings of the IEEE/CVF Conference on Computer Vision and Pattern Recognition}, 2022, pp. 5480--5490.

\bibitem{pauwels2015technicalofcbct}
R.~Pauwels \emph{et~al.}, ``Technical aspects of dental cbct: state of the art,'' \emph{Dentomaxillofacial Radiology}, vol.~44, no.~1, p. 20140224, 2015.

\bibitem{devereux2011cepImportant}
L.~Devereux \emph{et~al.}, ``How important are lateral cephalometric radiographs in orthodontic treatment planning?'' \emph{American Journal of Orthodontics and Dentofacial Orthopedics}, vol. 139, no.~2, pp. e175--e181, 2011.

\bibitem{deng2022depthnerf}
K.~Deng \emph{et~al.}, ``Depth-supervised nerf: Fewer views and faster training for free,'' in \emph{Proceedings of the IEEE/CVF Conference on Computer Vision and Pattern Recognition}, 2022, pp. 12\,882--12\,891.

\bibitem{yuan2023sparsenerfrgbd}
Y.-J. Yuan \emph{et~al.}, ``Neural radiance fields from sparse rgb-d images for high-quality view synthesis,'' \emph{IEEE Transactions on Pattern Analysis and Machine Intelligence}, vol.~45, no.~7, pp. 8713--8728, July 2023.

\bibitem{schwarz2020graf}
K.~Schwarz \emph{et~al.}, ``Graf: Generative radiance fields for 3d-aware image synthesis,'' \emph{Advances in Neural Information Processing Systems}, vol.~33, pp. 20\,154--20\,166, 2020.

\bibitem{jang2021codenerf}
W.~Jang and L.~Agapito, ``Codenerf: Disentangled neural radiance fields for object categories,'' in \emph{Proceedings of the IEEE/CVF International Conference on Computer Vision}, 2021, pp. 12\,949--12\,958.

\bibitem{peng2023codebody}
S.~Peng \emph{et~al.}, ``Implicit neural representations with structured latent codes for human body modeling,'' \emph{IEEE Transactions on Pattern Analysis and Machine Intelligence}, vol.~45, no.~8, pp. 9895--9907, Aug 2023.

\bibitem{garbin2021fastnerf}
S.~J. Garbin \emph{et~al.}, ``Fastnerf: High-fidelity neural rendering at 200fps,'' in \emph{Proceedings of the IEEE/CVF International Conference on Computer Vision}, 2021, pp. 14\,346--14\,355.

\bibitem{wadhwani2022squeezenerf}
K.~Wadhwani and T.~Kojima, ``Squeezenerf: Further factorized fastnerf for memory-efficient inference,'' in \emph{Proceedings of the IEEE/CVF Conference on Computer Vision and Pattern Recognition}, 2022, pp. 2717--2725.

\bibitem{zhang2020nerfpp}
K.~Zhang \emph{et~al.}, ``Nerf++: Analyzing and improving neural radiance fields,'' \emph{arXiv:2010.07492}, 2020.

\bibitem{Marschner2015fundamentalOfCG}
S.~Marschner \emph{et~al.}, \emph{\BIBforeignlanguage{English (US)}{Fundamentals of computer graphics, fourth edition}}.\hskip 1em plus 0.5em minus 0.4em\relax CRC Press, January 2015.

\bibitem{mescheder2019occupancynetworks}
L.~Mescheder \emph{et~al.}, ``Occupancy networks: Learning 3d reconstruction in function space,'' in \emph{Proceedings of the IEEE/CVF conference on computer vision and pattern recognition}, 2019, pp. 4460--4470.

\bibitem{sitzmann2019srn}
V.~Sitzmann \emph{et~al.}, ``Scene representation networks: Continuous 3d-structure-aware neural scene representations,'' in \emph{Advances in Neural Information Processing Systems}, vol.~32, 2019, pp. 1121--1132.

\bibitem{mildenhall2021nerf}
B.~Mildenhall \emph{et~al.}, ``Nerf: Representing scenes as neural radiance fields for view synthesis,'' \emph{Commun. ACM}, vol.~65, no.~1, p. 99–106, dec 2021.

\bibitem{muller2022instantngp}
T.~M\"uller \emph{et~al.}, ``Instant neural graphics primitives with a multiresolution hash encoding,'' \emph{ACM Transactions on Graphics}, vol.~41, no.~4, pp. 102:1--102:15, Jul. 2022.

\bibitem{liu2020nsvf}
L.~Liu \emph{et~al.}, ``Neural sparse voxel fields,'' \emph{Advances in Neural Information Processing Systems}, vol.~33, pp. 15\,651--15\,663, 2020.

\bibitem{chen2022tensorf}
A.~Chen \emph{et~al.}, ``Tensorf: Tensorial radiance fields,'' in \emph{European Conference on Computer Vision}, 2022, pp. 333--350.

\bibitem{yang2023freenerf}
J.~Yang \emph{et~al.}, ``Freenerf: Improving few-shot neural rendering with free frequency regularization,'' in \emph{Proceedings of the IEEE/CVF Conference on Computer Vision and Pattern Recognition}, 2023, pp. 8254--8263.

\bibitem{zhu2024mimlp}
H.~Zhu \emph{et~al.}, ``Is vanilla mlp in neural radiance field enough for few-shot view synthesis?'' \emph{arXiv preprint arXiv:2403.06092}, 2024.

\bibitem{kim2022infonerf}
M.~Kim \emph{et~al.}, ``Infonerf: Ray entropy minimization for few-shot neural volume rendering,'' in \emph{Proceedings of the IEEE/CVF Conference on Computer Vision and Pattern Recognition}, June 2022, pp. 12\,902--12\,911.

\bibitem{jain2021dietnerf}
A.~Jain \emph{et~al.}, ``Putting nerf on a diet: Semantically consistent few-shot view synthesis,'' in \emph{Proceedings of the IEEE/CVF International Conference on Computer Vision}, Oct 2021, pp. 5865--5874.

\bibitem{wynn2023diffusionerf}
J.~Wynn and D.~Turmukhambetov, ``Diffusionerf: Regularizing neural radiance fields with denoising diffusion models,'' in \emph{Proceedings of the IEEE/CVF Conference on Computer Vision and Pattern Recognition}, 2023.

\bibitem{zhou2023sparsefusion}
Z.~Zhou and S.~Tulsiani, ``Sparsefusion: Distilling view-conditioned diffusion for 3d reconstruction,'' in \emph{2023 IEEE/CVF Conference on Computer Vision and Pattern Recognition (CVPR)}, June 2023, pp. 12\,588--12\,597.

\bibitem{deng2023nerdi}
C.~Deng \emph{et~al.}, ``Nerdi: Single-view nerf synthesis with language-guided diffusion as general image priors,'' in \emph{Proceedings of the IEEE/CVF Conference on Computer Vision and Pattern Recognition (CVPR)}, June 2023, pp. 20\,637--20\,647.

\bibitem{xu2023neurallift}
D.~Xu \emph{et~al.}, ``Neurallift-360: Lifting an in-the-wild 2d photo to a 3d object with 360deg views,'' in \emph{Proceedings of the IEEE/CVF Conference on Computer Vision and Pattern Recognition (CVPR)}, June 2023, pp. 4479--4489.

\bibitem{GORDON1970art}
R.~Gordon \emph{et~al.}, ``Algebraic reconstruction techniques (art) for three-dimensional electron microscopy and x-ray photography,'' \emph{Journal of Theoretical Biology}, vol.~29, no.~3, pp. 471--481, 1970.

\bibitem{Mahmoud2019improvedTV}
G.~Mahmoudi \emph{et~al.}, ``Sparse-view statistical image reconstruction with improved total variation regularization for x-ray micro-ct imaging,'' \emph{Journal of Instrumentation}, vol.~14, pp. P08\,023--P08\,023, 08 2019.

\bibitem{Xu2020Schatten}
C.~Xu \emph{et~al.}, ``Sparse-view cbct reconstruction via weighted schatten p-norm minimization,'' \emph{Optics Express}, vol.~28, pp. 35\,469--35\,482, 11 2020.

\bibitem{Jiang2022dl}
B.~Jiang \emph{et~al.}, ``Deep learning reconstruction shows better lung nodule detection for ultra–low-dose chest ct,'' \emph{Radiology}, pp. 1--12, 01 2022.

\bibitem{ye2018deep}
D.~H. Ye \emph{et~al.}, ``Deep back projection for sparse-view ct reconstruction,'' in \emph{IEEE Global Conference on Signal and Information Processing}, Nov. 2018, pp. 1--5.

\bibitem{CHAO2022536}
L.~Chao \emph{et~al.}, ``Sparse-view cone beam ct reconstruction using dual cnns in projection domain and image domain,'' \emph{Neurocomputing}, vol. 493, pp. 536--547, 2022.

\bibitem{zhang2018densenet}
Z.~Zhang \emph{et~al.}, ``A sparse-view ct reconstruction method based on combination of densenet and deconvolution,'' \emph{IEEE Transactions on Medical Imaging}, vol.~37, no.~6, pp. 1407--1417, 2018.

\bibitem{han2018framingunet}
Y.~Han and J.~C. Ye, ``Framing u-net via deep convolutional framelets: Application to sparse-view ct,'' \emph{IEEE Transactions on Medical Imaging}, vol.~37, no.~6, pp. 1418--1429, 2018.

\bibitem{jin2017fbpconvnet}
K.~H. Jin \emph{et~al.}, ``Deep convolutional neural network for inverse problems in imaging,'' \emph{IEEE Transactions on Image Processing}, vol.~26, no.~9, pp. 4509--4522, Sep. 2017.

\bibitem{liao2018gans}
H.~Liao \emph{et~al.}, ``Adversarial sparse-view cbct artifact reduction,'' in \emph{Medical Image Computing and Computer Assisted Intervention}, 2018, p. 154–162.

\bibitem{lee2019dnn}
H.~Lee \emph{et~al.}, ``Deep-neural-network-based sinogram synthesis for sparse-view ct image reconstruction,'' \emph{IEEE Transactions on Radiation and Plasma Medical Sciences}, vol.~3, no.~2, pp. 109--119, 2019.

\bibitem{chung2023diffmbir}
H.~Chung \emph{et~al.}, ``Solving 3d inverse problems using pre-trained 2d diffusion models,'' in \emph{Proceedings of the IEEE/CVF Conference on Computer Vision and Pattern Recognition}, 2023, pp. 22\,542--22\,551.

\bibitem{lee2023perpdiff}
S.~Lee \emph{et~al.}, ``Improving 3d imaging with pre-trained perpendicular 2d diffusion models,'' in \emph{Proceedings of the IEEE/CVF International Conference on Computer Vision}, October 2023.

\bibitem{scarfe2008whatiscbct}
W.~C. Scarfe and A.~G. Farman, ``What is cone-beam ct and how does it work?'' \emph{Dental Clinics of North America}, vol.~52, no.~4, pp. 707--730, 2008.

\bibitem{Shen2019PatRecon}
L.~Shen \emph{et~al.}, ``Patient-specific reconstruction of volumetric computed tomography images from a single projection view via deep learning,'' \emph{Nature Biomedical Engineering}, vol.~3, pp. 880--888, 11 2019.

\bibitem{chen2021transunet}
J.~Chen \emph{et~al.}, ``Transunet: Transformers make strong encoders for medical image segmentation,'' \emph{arXiv:2102.04306}, 2021.

\bibitem{armato2011lidc}
S.~G. Armato \emph{et~al.}, ``The lung image database consortium (lidc) and image database resource initiative (idri): A completed reference database of lung nodules on ct scans,'' \emph{Medical Physics}, vol.~38, no.~2, p. 915–931, January 2011.

\end{thebibliography}

\end{document}